%% file: sample7.tex
\documentclass[trackchanges,twocolumn]{aastex7}
\usepackage{caption}
\usepackage{graphicx}
\usepackage{adjustbox}

\newcommand{\omatter}{\ensuremath{\Omega_{\mathrm{M}}}}

\newcommand{\caiihk}{\ensuremath{\mathrm{Ca}\,\textsc{ii}\,\mathrm{H\&K}}}

\newcommand{\si}{\ensuremath{\mathrm{Si}\,\textsc{ii}\,\lambda6355}}
\newcommand{\mg}{\ensuremath{\mathrm{Mg}\,\textsc{ii}}}
\newcommand{\cahk}{\ensuremath{\mathrm{Ca}\,\textsc{ii\,H\&K}}}

\begin{document}

\title{What Shapes the Ultraviolet Spectra of Type Ia Supernovae?}

\author[orcid=0000-0002-7350-7043]{Snehasish~Bhattacharjee}
\affiliation{Graduate Institute of Astronomy, National Central University, 300 Zhongda Road, 32001 Zhongli, Taiwan}
\email[show]{snehasish@astro.ncu.edu.tw}  

\author[orcid=0000-0001-8415-6720]{Yen-Chen~Pan}
\affiliation{Graduate Institute of Astronomy, National Central University, 300 Zhongda Road, 32001 Zhongli, Taiwan}
\email[show]{ycpan@astro.ncu.edu.tw}  

\author[orcid=0000-0002-7708-3831]{Anirban~Dutta}
\affiliation{Graduate Institute of Astronomy, National Central University, 300 Zhongda Road, 32001 Zhongli, Taiwan}
\email{anirbandutta@gm.astro.ncu.edu.tw} 

\author[orcid=0009-0005-3795-4865]{Shubham~Gupta}
\affiliation{Graduate Institute of Astronomy, National Central University, 300 Zhongda Road, 32001 Zhongli, Taiwan}
\email{d1109602@gm.astro.ncu.edu.tw} 

\author[orcid=0009-0002-5837-2504]{Chia-Yu~Cheng}
\affiliation{Graduate Institute of Astronomy, National Central University, 300 Zhongda Road, 32001 Zhongli, Taiwan}
\email{m1129003@gm.astro.ncu.edu.tw} 

\author[orcid=0000-0002-5740-7747]{Charles D. Kilpatrick}
\affiliation{Center for Interdisciplinary Exploration and Research in Astrophysics (CIERA), Northwestern University, Evanston, IL 60201, USA}
\email{ckilpatrick@northwestern.edu} 

\author[orcid=0000-0003-2445-3891]{Matthew~R.~Siebert}
\affiliation{Space Telescope Science Institute, Baltimore, MD 21218, USA}
\email{msiebert@stsci.edu}

\begin{abstract}

Ultraviolet (UV) observations of Type~Ia supernovae (SNe~Ia) exhibit significant diversity, but their physical origin remains poorly understood. We investigate UV spectra of SNe~Ia observed with the {\it Swift} UltraViolet and Optical Telescope. Our expanded pre-maximum sample probes dependencies on decline rate, ejecta properties, and host environment. We recover the correlation between UV flux and decline rate near maximum in the near-UV (NUV), and find correlations at pre-maximum phases in both the NUV and mid-UV (MUV). This trend is strongly wavelength dependent, with slow decliners showing higher NUV but lower MUV flux than fast decliners. Synthetic \textsc{tardis} spectra varying the $^{56}$Ni abundance show that enhanced $^{56}$Ni suppresses the MUV while increasing the NUV, broadly reproducing the trend. We find no significant correlation between UV flux and ejecta velocity, while Si~II $\lambda6355$ pEW shows trends similar to those with decline rate. We also find no significant correlations with host-galaxy properties, although host environment may contribute to the remaining UV diversity among SNe with similar decline rates. Early-excess SNe show enhanced flux mainly in the NUV rather than a broad UV excess, suggesting that strong circumstellar interaction is not necessarily required to explain the observed excess emission. Peculiar subclasses show distinct UV properties, with brighter and hotter subclasses showing higher UV flux, although the current sample is small and does not cover all SN~Ia subtypes. Our results suggest that the UV diversity of SNe~Ia is shaped by multiple mechanisms that contribute differently across the UV wavelength range.

\end{abstract}



\section{Introduction}
Type Ia supernovae (SNe~Ia) are one of the most luminous astrophysical transients, often outshining their host galaxies, with bolometric luminosities of the order of $10^{43} \, \text{erg s}^{-1}$. They play a crucial role in astrophysics, as they produce a significant portion of the intermediate-mass and iron-group elements (IGEs) in the Universe, which contribute substantially to the chemical evolution of galaxies \citep[e.g.,][]{nomoto1984,iwamoto1999,matteucci2009,kobayashi2009,lach2020}. Furthermore, they show a correlation between their peak luminosities and the post-peak decline rates \citep[e.g.,][]{phillips1993,phillips1999,kattner2012}. This allows them to be used as standardizable candles to measure extragalactic distances, leading to the discovery of accelerated expansion of the Universe \citep{reiss1998,perlmutter1999}. 

While SNe~Ia are thought to result from thermonuclear explosions of white dwarfs (WDs) in binary systems, questions remain regarding their progenitor scenario(s) and explosion mechanism(s). Ultraviolet (UV) spectra of SNe~Ia contain a wealth of information regarding their outermost layers \citep[e.g.,][]{hachinger2013,cao2015}, which depend on both the properties of the progenitor system and explosion \citep[e.g.,][]{sauer2008,brown2015,foley2016,derKacy2020,derKacy2023}. Previous studies have revealed substantial UV spectral diversity in SNe~Ia. For instance, \citet{sullivan2009} found that higher redshift SNe~Ia show weaker \si, \caiihk, and \mg\ UV features, implying lower intermediate-mass element (IME) abundances for these higher redshift events. 
\citet{ellis2008} showed that the rest-frame UV spectra of SNe~Ia at intermediate redshifts exhibit significant diversity, while \citet{cooke2011} found that the spectral properties of a sample of nearby SNe~Ia are largely consistent with those at intermediate redshifts, with only minor differences in metallic absorption lines, and suggested the variations are likely driven by compositional differences and not due to evolution. \citet{wang2012} reported systematic differences in the near-UV (NUV) spectra that cannot be attributed to progenitor metallicity or ejecta velocities, and instead suggested that viewing angle effects arising from asymmetric explosions may play a role. \citet{Maguire2012} found that low-redshift SNe~Ia show suppressed NUV flux compared to intermediate-redshift ones, possibly due to metallicity differences, and identified a correlation between light-curve width and the velocities of NUV spectral features, and argued that the host galaxy properties could influence these trends. Using {\it Swift} photometry, \citet{milne2013} revealed distinct UV color groups in SNe~Ia, with NUV-blue events showing less iron absorption, unburned carbon, and lower \si\ velocity gradients than their NUV-red counterparts. Recently, \citet{bhattacharjee2025} found that SN~2003fg-like events (03fg-like; e.g., \citealt{Howell2006,scalzo2010}) exhibit unusually strong UV flux at all phases compared to normal SNe~Ia, where the observed UV excess cannot be explained by host galaxy properties and instead could be attributed to the presence of an additional hot blackbody component, possibly arising from the interactions with nearby circumstellar material (CSM). 

Theoretical studies predict that the UV flux of SNe~Ia is influenced by the metallicity of their progenitors \citep{lentz2000, walker2012}, which is consistent with observational studies that report SNe~Ia residing in metal-poor hosts to show higher UV flux \citep{pan2020}, though the claim is debated in \citet{brown2020}. Furthermore, UV observations have also been suggested to aid in luminosity calibrations and enhance the reliability of SNe~Ia as cosmological standard candles \citep[e.g.,][]{foley2008,brown2010}. These findings, therefore, highlight the importance of UV observations in probing the diversity and underlying physics of SNe~Ia, motivating a comprehensive analysis using a large and homogeneous dataset.

In this work, we analyze the UV spectra of $34$ nearby ($z\lesssim 0.02$) SNe~Ia from the {\it Swift} UltraViolet and Optical Telescope (UVOT) grism observations. We investigate relationships between UV spectral properties and $\Delta m_{15}(B)$\footnote{$\Delta m_{15}(B)$ represents the decline rate of SN~Ia light curve, defined as the decline in $B$-band magnitude between peak and $15$\,days after peak in the rest-frame B-band light curve (see also \citealt{phillips1993}).}, ejecta characteristics, and host-galaxy properties. We also compare the UV spectra of peculiar SNe~Ia with their normal counterparts. The article is organized as follows: In Section~\ref{sec:sample}, we describe our sample and data. Section~\ref{sec:compare} presents the spectroscopic analysis. The discussion and conclusions are presented in Sections~\ref{sec:discuss} and \ref{sec:conclude}, respectively. Throughout this paper, we assume $\mathrm{H_0}=70$\,km\,s$^{-1}$\,Mpc$^{-1}$ and a flat universe with $\omatter=0.3$.

\section{Data}\label{sec:sample}

\subsection{SN~Ia sample}
\label{sec:normal-sample}

The UV spectral observations were obtained with the {\it Swift} UVOT grism \citep{gehrels2004,roming2004,kuin2015}. Due to the slitless design of the grism, the spectra are often contaminated by nearby background sources (e.g., their host galaxies). To correct for this, we employed the data reduction pipeline outlined in \citet{pan2018,pan2020} and further improved in \citet{bhattacharjee2025}. The pipeline allows users to construct a background spectrum by interactively selecting regions adjacent to the target spectrum, which is then subtracted to recover the intrinsic SN signal. 

Our sample comprises SNe~Ia that were previously investigated in \citet{pan2018,pan2020}, and \citet{bhattacharjee2025}, along with three newly observed events, namely SN~2024inv, SN~2025fvw, and SN~2025rbs, for a total of $34$ objects. The $\Delta m_{15}(B)$ and host-galaxy $E(B-V)$ are obtained by fitting their optical light curves using the \texttt{EBV\_model} in the \texttt{SNooPY} package \citep{burns2011}. The light curves and the \texttt{SNooPY} template fittings for newly added SNe are shown in appendix~\ref{snpy:app}. The optical light curves are obtained from the Lulin One-meter Telescope (LOT), Zwicky Transient Facility (ZTF), and {\it Swift}. The LOT photometry were processed using standard procedures, including bias subtraction, flat-field correction, and photometric calibration against field stars observed in the same frame. For {\it Swift} photometry, we retrieved the UVOT imaging data from the HEASARC archive and reduced them with \textsc{HEAsoft} \citep{heasarc2014} using the \texttt{uvotsource} task. Aperture photometry was extracted at the SN position with a source radius of $3\arcsec$--$5\arcsec$ and a nearby sky region free of contaminating sources, following the prescriptions developed for the {\it Swift} Optical/Ultraviolet Supernova Archive \citep[SOUSA;][]{brown2014sousa,brown2009} and recent multi-wavelength SN studies \citep[e.g.,][]{kilpatrick2018,jencson2024}. Count rates were converted to AB magnitudes using the UVOT photometric zeropoints and time-dependent sensitivity corrections of \citet{breeveld2010}. When host-galaxy contamination was non-negligible, pre-explosion template images were used to subtract the local background at the SN position and to apply the PSF-loss corrections described by \citet{brown2014sousa}.

\begin{figure}[htbp]
  \centering
  \includegraphics[width=0.45\textwidth]{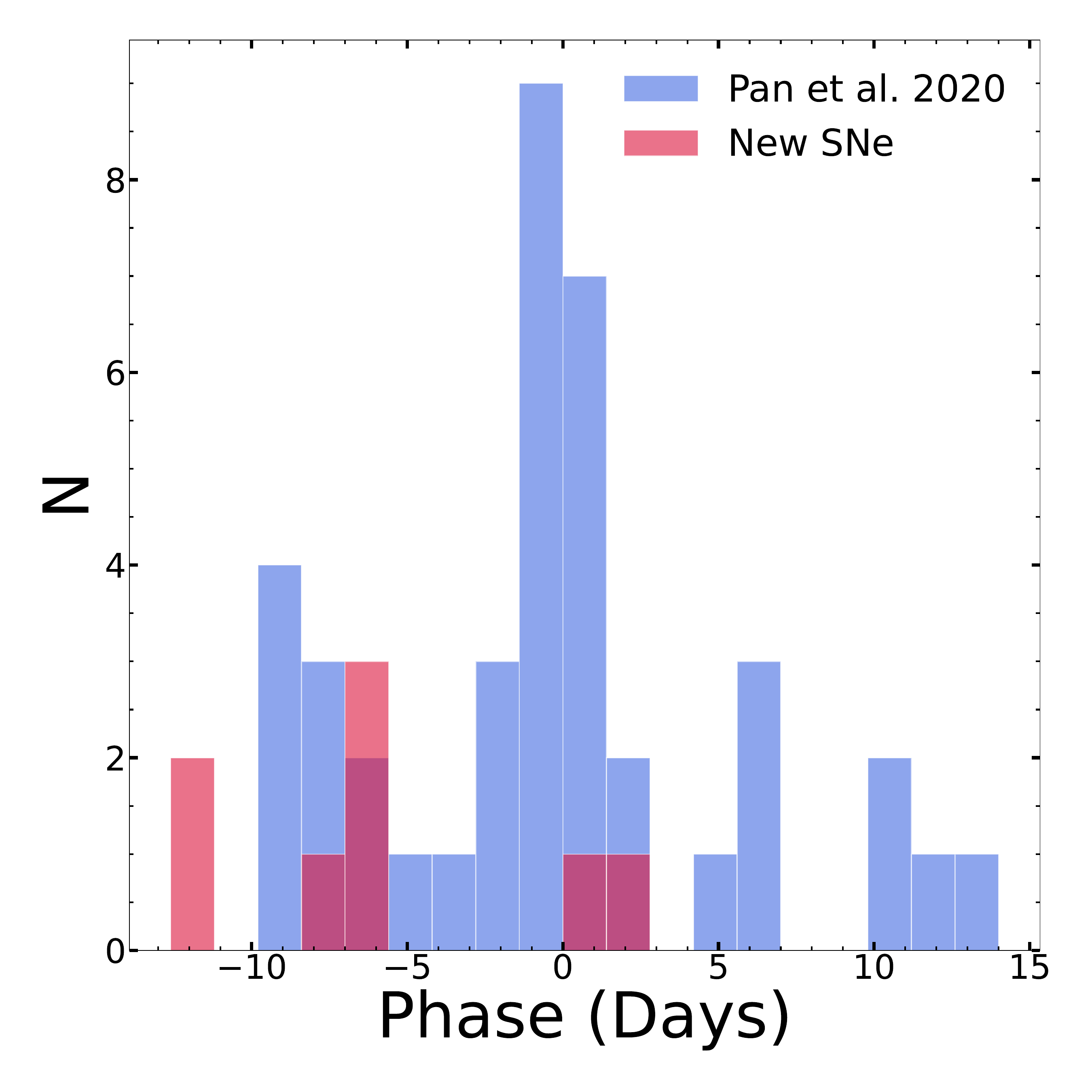}
\caption{Phase distribution of our SN~Ia sample. Blue histogram represents SNe previously studied in \citet{pan2020}, while red histogram indicates additional SNe included in this work.}
  \label{fig:phase_hist}
\end{figure}

Figure~\ref{fig:phase_hist} shows the phase distribution of our sample, where the blue bars represent SNe previously studied in \citet{pan2020}, while the red bars indicate additional SNe included in this work. The number of SNe with early-time observations (i.e., phase $<-5$~days) has increased from $9$ in \citet{pan2020} to $15$ in this work. Early-time UV spectra of SNe~Ia probe the outermost layers of the ejecta, and are highly sensitive to the progenitor composition and explosion physics, and provide valuable diagnostics to characterize the abundances of IGEs \citep{hachinger2013,milne2013}. The inclusion of more early-time data in this work offers improved insight into the UV diversity of SNe~Ia.

We correct the UV spectra for Milky-Way reddening using the calibrations of \citet{sf2011} and host-galaxy reddening obtained from the light curve template fitting. Here we use $R_{V} = 3.1$ and a \citet*[][CCM]{ccm89} reddening law. However, our results are insensitive to the choice of the reddening law, such as the \citet{fitzpatrick2007} prescription, or other \texttt{SNooPY} light-curve fitting models, producing only negligible changes in the derived quantities. A summary of the SNe~Ia sample used in this work is shown in Table~\ref{tab:sn_host_properties}. The final reduction of our UV spectra in this study is available on 
\texttt{Zenodo}.\footnote{\url{https://doi.org/10.5281/zenodo.22646347}}

\subsection{Host-galaxy properties}

We determine the host-galaxy properties of our sample to investigate their potential connection with the UV spectral properties. We use the \textsc{PYTHON} package \texttt{PROSPECTOR} \citep{johnson2019,johnson2021} based on a Bayesian forward modeling approach to fit the spectral energy distribution (SED) of the host galaxies with their multicolor photometry. The code utilizes the Flexible Stellar Population Synthesis package  \citep[\texttt{FSPS};][]{conroy2009} to obtain the SED models using \textsc{PYTHON-FSPS} \citep{foreman2014}. We obtain the best-fit SED using the \texttt{emcee} package \citep{foreman2013}, which is based on the Markov-Chain Monte Carlo (MCMC) sampling employing the Metropolis-Hastings algorithm \citep{metropolis1953,hastings1970}. From the best-fit SEDs, we derive the host-galaxy stellar mass, star-formation rate (SFR), specific star-formation rate (sSFR), stellar age, and stellar metallicity.

The host-galaxy photometry is obtained from the GALEX \citep{martin2005, morrissey2007}, Pan-STARRS~1 \citep[PS1;][]{chambers2016}, SDSS \citep{york2000}, 2MASS \citep{skrutskie2006}, and WISE \citep{wright2010} imaging data, with a wavelength coverage from UV to infrared (IR). The photometry is measured using the \textsc{PYTHON} package \texttt{HostPhot} \citep{hostphot2022}. The code applies an aperture chosen by visual inspection to the imaging data for each filter by using Kron flux parameters provided by the Python library for Source Extraction and Photometry \citep[\texttt{SEP};][]{Bertin_Arnouts_SEXtrator,Barbary2016_SEP}. The same aperture size is applied to all the images with the PS1 $g$-band image as the reference. The photometry is then corrected for foreground Milky Way extinction using the CCM reddening law with \( R_V = 3.1 \). Summary of the host-galaxy properties is shown in Table~\ref{tab:sn_host_properties}.

We also examined the availability of gas-phase metallicity measurements for the larger sample in this work. However, compared with \citet{pan2020}, only one additional host galaxy has emission-line measurements for deriving a gas-phase metallicity. For the remaining newly added objects, either no suitable host-galaxy spectrum is available or the host is passive and lacks sufficiently strong emission lines. Therefore, the gas-phase metallicity sample has not increased significantly, and we use the stellar metallicities derived from the host-galaxy SED fitting in the analysis.

\subsection{Ejecta properties}

We derive the ejecta properties of our sample, such as the ejecta velocities of \si\ ($v_{\si}$) and \cahk, and their pseudo-equivalent widths (pEWs), to examine their correlations with the UV spectral properties. We obtain the near-peak (within $\pm5$\,days of maximum light) optical spectra of our sample from the WISeREP database \citep{yaron2012}. The prescriptions used to measure the ejecta properties will be outlined in detail in an upcoming paper (S.~Gupta~et~al., in prep). In brief, the ejecta velocities are measured by fitting the absorption feature with multiple Gaussian profiles in wavelength space, where the center of each profile is determined by the expansion velocity through the Doppler shift. For blended features such as \cahk, we assume all transitions within a specific component to share a common velocity to ensure robust measurements against blending. The pEWs are then measured by estimating the total Gaussian area for each component separately.

\begin{table*}
\centering
\caption{Summary of the SN~Ia sample and their properties in this work. A portion is shown here for guidance regarding its form and content. The entire table is available online as supplementary material.} 
\label{tab:sn_host_properties}
\tiny
\setlength{\tabcolsep}{1pt}
\begin{tabular}{ccccccccccccc}
\hline
SN Name & Redshift & Phase & $\Delta m_{15}(B)^{a}$  & $E(B{-}V)_{\mathrm{host}}$ & $\log\text{M*}^{b}$ & $\log\text{Z}^{c}$ & $\text{Age}^{d}$ & $\log\text{SFR}^{e}$ & $\log\text{sSFR}^{f}$  & $v_{\si}^{g}$ & $\text{pEW}_{\si}^{h}$\\
 &  & (days) & (mag) &  (mag) & ($M_\odot$) & $(Z/ Z_\odot)$ & (Gyr) & $(M_\odot\,\mathrm{yr}^{-1})$ & ($\mathrm{yr}^{-1}$) & ($km \ s^{-1}$) &\,(\AA)\ \\
\hline
SN~1992A & $0.0062$  & $+4.9$ & $1.47 \pm 0.05$  & $0.00 \pm 0.02$ & $10.914^{+0.04}_{-0.033}$ & $+1.208^{+0.088}_{-0.081}$ & $2.078^{+0.104}_{-0.118}$ & $-1.022^{-1.052}_{-0.546}$ & $-11.947^{+11.496}_{-11.98}$ & $-11803 \pm 53$ & $113.73 \pm 1.66$ \\
SN~2005cf & $0.0065$ & $-6.9$ & $1.05 \pm 0.03$  & $0.10 \pm 0.03$ & $9.519^{+0.093}_{-0.080}$ & $-0.832^{+0.106}_{-0.109}$ & $3.892^{+1.742}_{-1.046}$ & $-0.602^{+1.833}_{-1.779}$ & $-10.121^{+10.66}_{-10.905}$  & $-10155 \pm 16$  & $97.58 \pm 0.47$ \\
SN~2005df & $0.0044$ & $+0.7,-6.1$ & $1.12 \pm 0.00$  & $0.04 \pm 0.01$ & $10.375^{+0.004}_{-0.004}$  & $+1.469^{+0.029}_{-0.024}$ & $2.359^{+0.036}_{-0.035}$ & $-5.564^{+4.744}_{-5.679}$ & $-15.938^{+15.119}_{-16.053}$ & ... & ... \\
SN~2005ke & $0.0049$ & $-2.7$ & $1.76 \pm 0.01$  & $0.00 \pm 0.02$ & $10.937^{+0.048}_{-0.048}$ & $-0.478^{+0.105}_{-0.096}$  & $10.754^{+1.967}_{-1.746}$ & $-11.073^{+3.212}_{-11.073}$ & $-22.012^{+14.132}_{-22.012}$ & $-10545 \pm 42$ & $97.32 \pm 1.13$ \\
SN~2008Q & $0.0079$ & $-0.2$ & $1.41 \pm 0.05$& $0.09 \pm 0.03$ & $11.513^{+0.013}_{-0.010}$ & $+0.043^{+0.014}_{-0.020}$ & $10.325^{+0.547}_{-0.383}$ & $-0.999^{+2.404}_{-2.395}$ & $-12.513^{+13.849}_{-13.777}$ & $-11069 \pm 24$  & $95.81 \pm 0.71$ \\
SN~2009an & $0.0092$ & $+2.7$ & $1.44 \pm 0.00$  & $0.09 \pm 0.03$ & $10.767^{+0.026}_{-0.019}$ & $+0.348^{+0.045}_{-0.056}$ & $6.404^{+0.607}_{-0.443}$ & $-0.767^{+1.793}_{-1.911}$ & $-11.534^{+11.534}_{-12.744}$ & ... & ... \\
\ldots & \ldots & \ldots & \ldots & \ldots & \ldots & \ldots & \ldots & \ldots & \ldots & \ldots & \ldots &  \\

\ldots & \ldots & \ldots & \ldots & \ldots & \ldots & \ldots & \ldots & \ldots & \ldots & \ldots & \ldots &  \\

\ldots & \ldots & \ldots & \ldots & \ldots & \ldots & \ldots & \ldots & \ldots & \ldots & \ldots & \ldots & \\

SN~2024inv & $0.0045$ & $+1.1$ & $1.04 \pm 0.09$ & $0.00 \pm 0.08$ & $10.002^{+0.078}_{-0.058}$ & $-0.265^{+0.141}_{-0.161}$ & $5.322^{+1.967}_{-1.347}$ & $-4.647^{+2.077}_{-4.647}$ & $-14.645^{+12.093}_{-14.645}$ & ... & ... \\
SN~2025fvw & $0.0061$ & $-6.0$ & $1.20 \pm 0.02$  & $0.08 \pm 0.01$ & $10.444^{+0.011}_{-0.019}$ & $-0.534^{+0.048}_{-0.039}$ & $13.056^{+0.555}_{-1.036}$ & $-0.190^{+1.589}_{-1.546}$ & $-10.632^{+11.889}_{-11.926}$ & $-9675 \pm 456$ & $60.07 \pm 15.52$ \\
SN~2025rbs & $0.0027$ & $+2.4$ & $1.50 \pm 0.02$ 1 & $0.00 \pm 0.02$ & $10.894^{+0.001}_{-0.001}$ & $+1.052^{+0.005}_{-0.006}$ & $13.757^{+0.032}_{-0.071}$ & $+0.222^{+1.743}_{-1.747}$ & $-10.673^{+12.536}_{-12.563}$ & ... & ... \\
\hline
\end{tabular}
\begin{flushleft}
$^a${The decline in $B$-band magnitude 15~days after the peak brightness.}\\
$^b${Host-galaxy stellar mass.}\\
$^c${Host-galaxy stellar metallicity.}\\
$^d${Host-galaxy stellar age.}\\
$^e${Host-galaxy star formation rate.}\\
$^f${Host-galaxy specific star formation rate.}\\
$^g${Photospheric velocity of the \si\ line.}\\
$^h${Photospheric pEW of the \si\ line.}
\end{flushleft}
\end{table*}

\section{Results}\label{sec:compare}

\subsection{Mean spectra}\label{sec:mean}

Analysis with mean UV spectra has been reported to be invaluable in examining trends with various parameters of SNe~Ia, such as their $\Delta m_{15}(B)$, host-galaxy stellar mass ($M_{*}$), and metallicity ($Z$) \citep[e.g.,][]{foley2016,pan2020,bhattacharjee2025}. In this work, we further extend the analysis to include host-galaxy specific star-formation rate (sSFR), stellar age, and velocities and pEWs of \si\ and \cahk\ features. We generate mean spectra by first normalizing each spectrum to its median flux in the wavelength range $4000–4500$\,\AA, followed by interpolation onto a common wavelength grid. To examine trends with different parameters, the sample is then divided into two subsamples based on the threshold values of the parameters, and the mean spectrum of each subsample is obtained by averaging over the normalized fluxes. Uncertainties on the mean spectra are estimated using the bias-corrected and accelerated (BCa) bootstrapping method \citep{efron1987,efron1993}. Results are shown in Figure~\ref{fig:mean:spec}. The peculiar SNe~Ia have been excluded from this analysis as their distinct spectral properties could produce biased trends, and therefore only spectroscopically normal SNe~Ia events are considered. We note that the analysis presented here is qualitative, with their statistical significance assessed separately in Section~\ref{section:flux-ratio}.

We define two spectroscopic subsamples for our analysis. The near-peak sample consists of SNe~Ia observed between $-5$\,days and $+5$\,days relative to $B$-band maximum, with a median phase of $-0.1$\,day, and are shown in red, while the pre-maximum sample includes SNe~Ia observed between $-10$\,days and $0$\,days relative to $B$-band maximum, with a median phase of $-5.7$\, day, and shown in blue. This pre-maximum phase cut is motivated by previous studies showing that the UV light curves of SNe~Ia typically peak several days before the optical maximum \citep[e.g.,][]{Jha2006,Milne2010}, and is therefore chosen so that the mean epoch of the pre-maximum sample is around the expected UV maximum. 

We begin by examining the relationship with $\Delta m_{15}(B)$ in panel (a) of Figure~\ref{fig:mean:spec}. Previous studies \citep[e.g.,][]{foley2016,pan2020} have shown that near-peak SNe~Ia with higher $\Delta m_{15}(B)$ tend to exhibit reduced UV flux in the region of $\sim 3000$\,\AA. To investigate this trend using our larger near-peak and pre-maximum samples, we further separate each sample into two sub-groups using the median value of $\Delta m_{15}(B)$. The bottom panels show the ratio spectrum of the two sub-groups, from which it is clear that slower-declining (with lower $\Delta m_{15}(B)$) SNe~Ia tend to exhibit higher UV flux between $2800$ and $3500$\,\AA\ with the near-peak sample. In contrast, we observe that fast-declining (with higher $\Delta m_{15}(B)$) SNe~Ia tend to show higher flux at wavelengths $\lesssim 2800$\,\AA\ in both our pre-maximum and near-peak samples. We also note that the slower decliners tend to present more significant absorption features at $\sim3200$\,\AA\ than those of fast decliners.

Panel (b) of Figure~\ref{fig:mean:spec} illustrates the result for host-galaxy stellar metallicity. We divide the sample into metal-poor and metal-rich sub-groups using the median stellar metallicity. \citet{pan2020} reported that near-peak SNe~Ia in more metal-poor host galaxies tend to exhibit higher UV flux levels at wavelengths $\lesssim 2800$\,\AA. However, we do not find any clear trend with stellar metallicity in either the near-peak or pre-maximum sample. Note that we adopt the stellar metallicity rather than gas-phase metallicity to maximize our sample size since the host-galaxy spectroscopic data are not available for all of the SNe in our sample, and not all of the host galaxies are star-forming, preventing reliable gas-phase metallicity measurements. Panel (c) shows the trend for host-galaxy stellar mass $M_{*}$. Following \citet{pan2020}, we divide the sample into two groups at $\log(M_{*}/M_{\odot}) = 10$. We observe no clear difference between the UV flux levels of the low- and high-mass sub-groups for either the pre-maximum or near-peak sample. Panels (d) and (e) present the results for host-galaxy stellar age and sSFR. The samples are divided into two sub-groups using the median value of each parameter. No clear trend is found with either the host-galaxy stellar age or sSFR.

Ejecta properties of SNe~Ia, such as $v_{\si}$ and their pEWs provide important diagnostics of explosion physics and progenitor systems. For instance, near-peak \si\ velocities are often used to further classify normal SNe~Ia into high-velocity (HV SNe~Ia; $v_{\si}$ $\geq 12{,}000$ km $s^{-1}$) and normal-velocity (NV SNe~Ia; $v_{\si}$ $< 12{,}000$ km $s^{-1}$) sub-groups, which can improve the standardization of SNe~Ia \citep[e.g.,][]{wang2009,Siebert2020}. The pEWs of these lines probe the ionization states and are key to understanding both the diversity and luminosity calibration of SNe~Ia. Panel (f) presents the results for $v_{\si}$. We divide the sample at the median velocity of the sample rather than using $12{,}000~\mathrm{km~s^{-1}}$ threshold, since only $5$ SNe in our sample have $v_{\si} \geq 12{,}000~\mathrm{km~s^{-1}}$. A mean spectrum constructed from such a small number of SNe would not be representative. As expected, the mean spectrum of the higher-velocity sub-group shows more blueshifted UV spectral features than that of the lower-velocity sub-group. We also find that SNe~Ia with higher $v_{\si}$ generally show lower UV flux levels than those with lower $v_{\si}$ over the wavelength range of $\sim3000$--$3500$\,\AA. In contrast, the higher-velocity sub-group appears to show a higher flux level below $\sim2800$\,\AA. However, this difference is not statistically significant based on our quantitative analysis in Section~\ref{section:flux-ratio}.

Finally, we investigate the relationship with the pEW of the \si\ absorption feature in panel (g). The sample is divided into two groups based on the median pEW values. For the pre-maximum sample, we do not find any significant trend. In contrast, for the near-peak sample, we find that the mean spectrum with smaller pEW tends to show a higher UV flux level in the $\sim 2800–3500$\,\AA\ range, similar to the trend found for ejecta velocity. 

We also construct mean spectra with the \cahk\ velocity and pEW measurements, and find the trends are generally consistent with those of \si. These results are shown in Appendix~\ref{cahk:app}.

\begin{figure*}[htbp]
  \centering
  \includegraphics[width=0.79\textwidth]{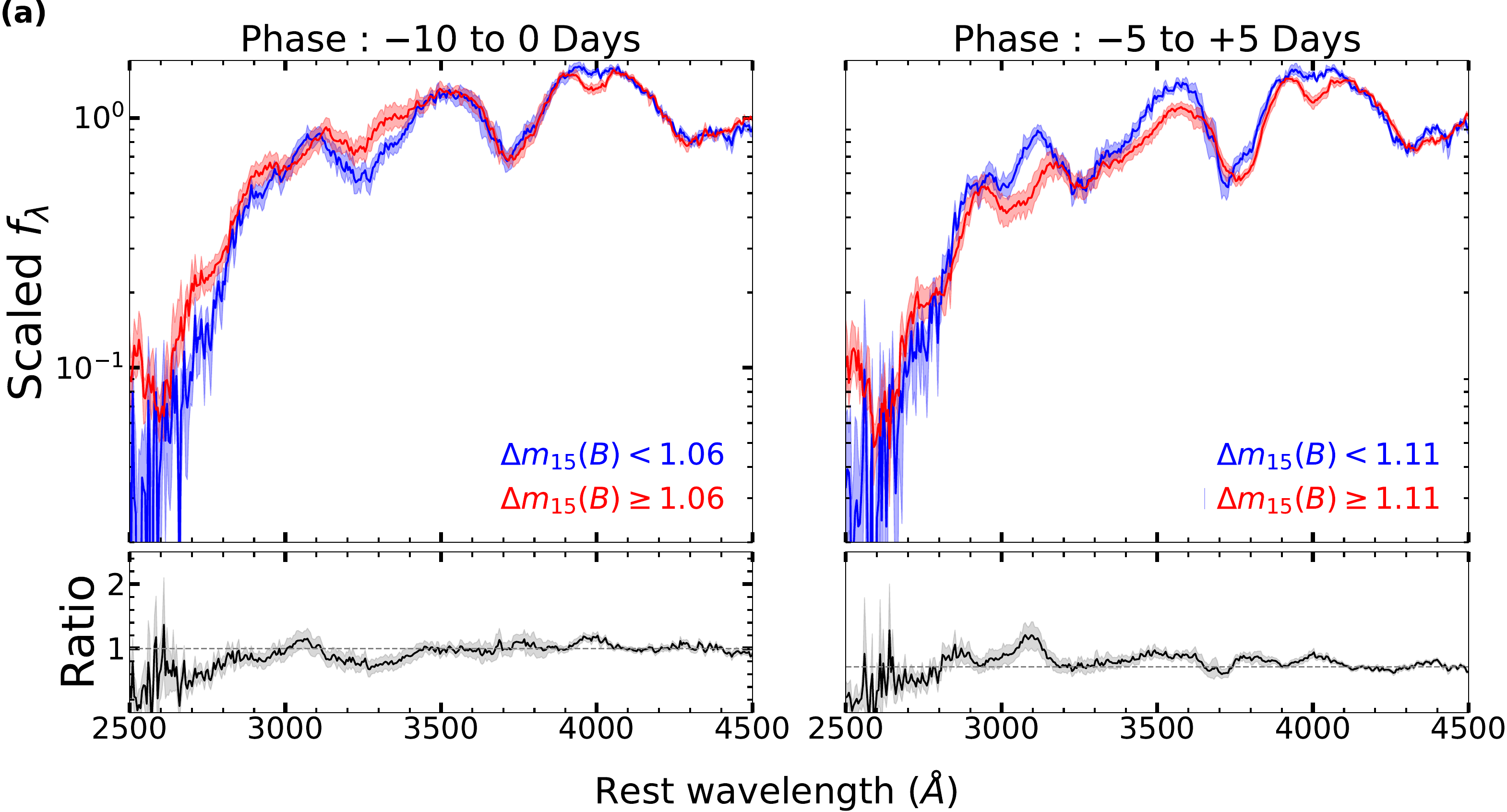}
  \includegraphics[width=0.79\textwidth]{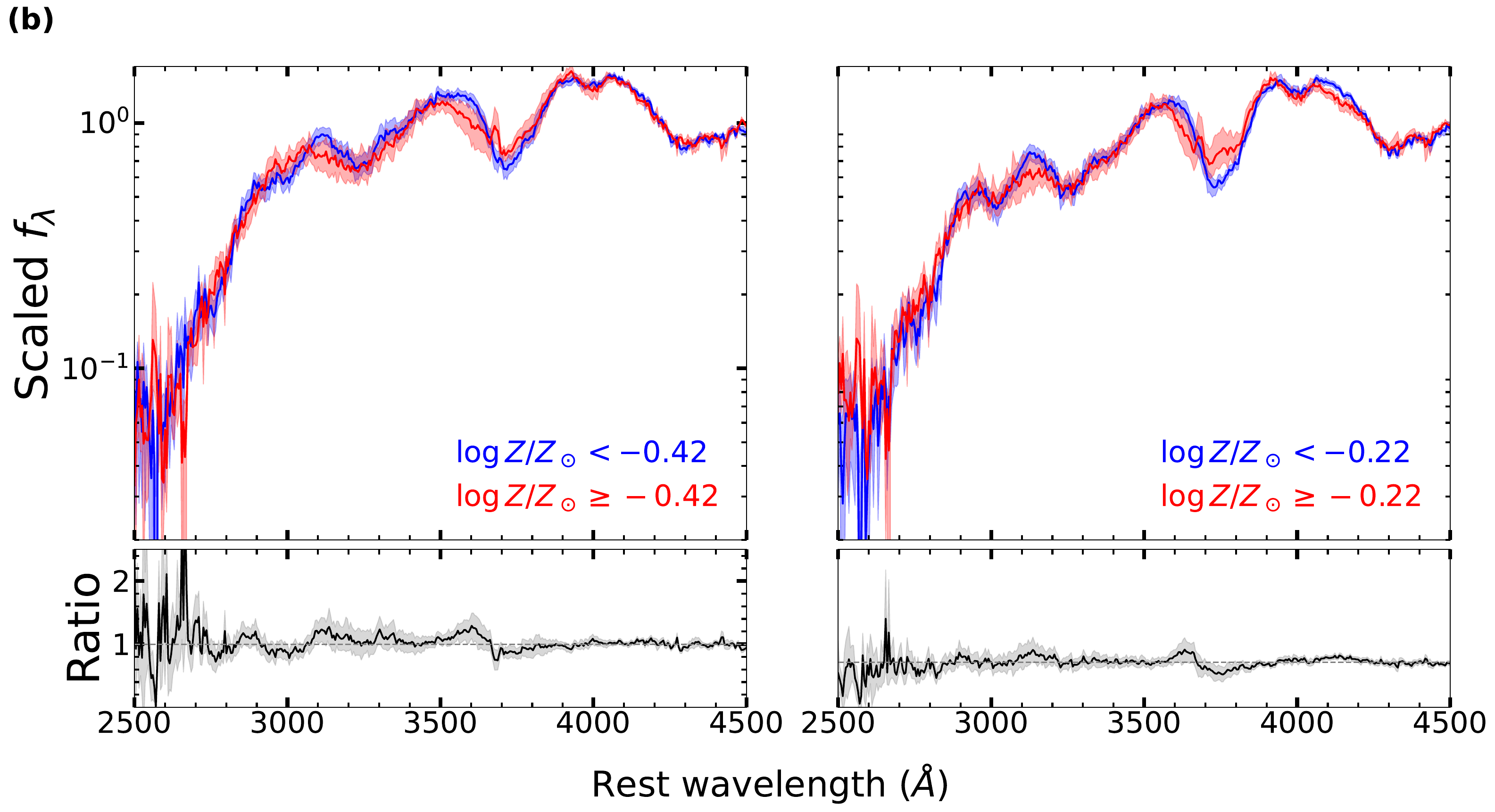}
  \includegraphics[width=0.79\textwidth]{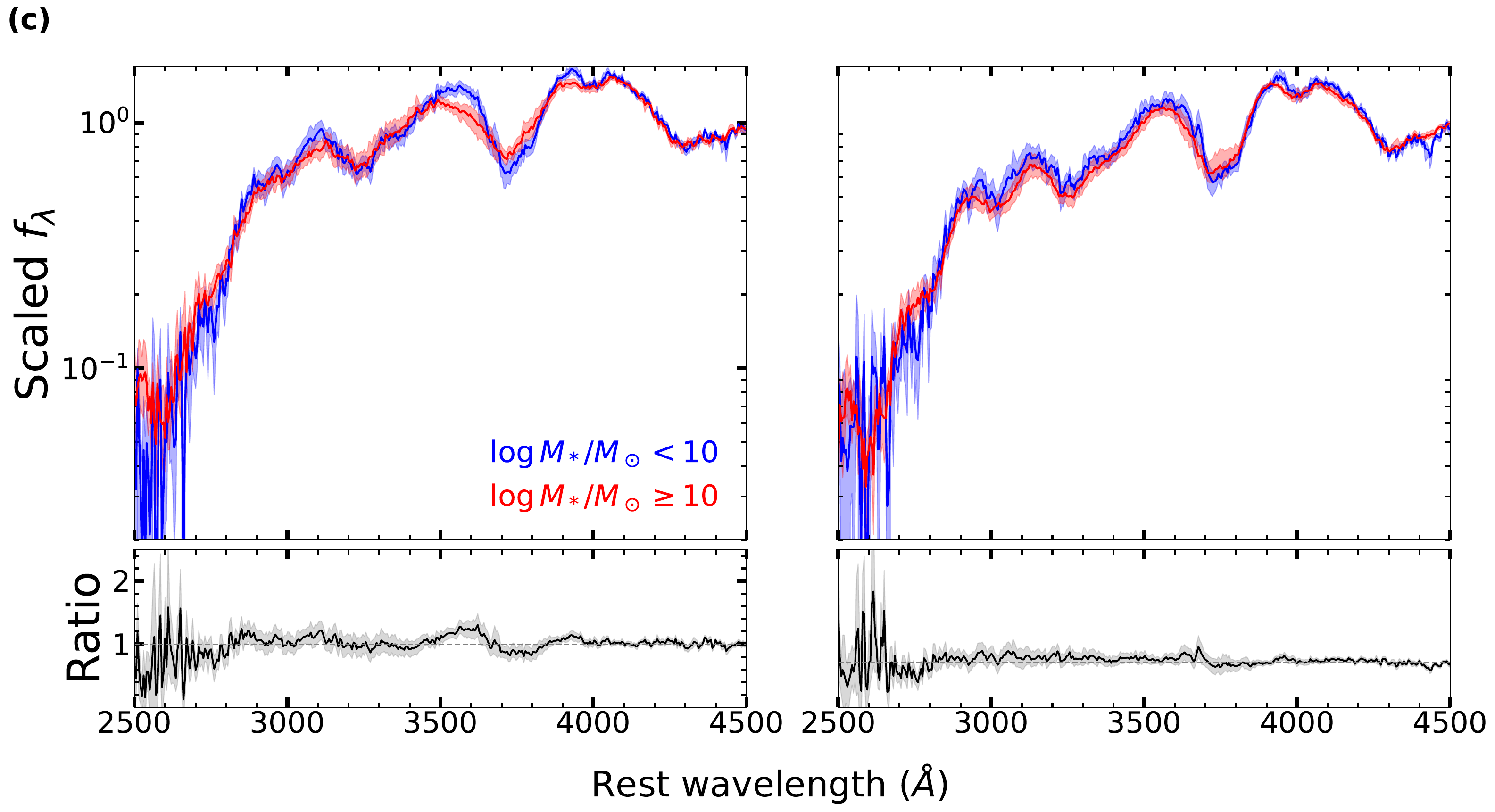}
  \caption{(a) \textit{Left}: Mean spectra of {\it Swift} pre-maximum sample (i.e., $-10$\,d $\leq$ phase $\leq$ $ 0$\,d) with respect to $\Delta m_{15}(B)$. The bottom panel shows the ratio of the mean spectra for SNe with  $\Delta m_{15}(B) < 1.06$ to those with $\Delta m_{15}(B) \geq 1.06$. \textit{Right}: Same as the left but for near-peak sample (i.e., $-5$\,d $\leq$ phase $\leq$ $ +5$\,d). (b) Same as (a) but with respect to host-galaxy metallicity $Z$. (c) Same as (a) but with respect to host-galaxy stellar mass $M_{*}$.}
  \label{fig:mean:spec}

\end{figure*}

\begin{figure*}[htbp]\ContinuedFloat
  \centering
  \includegraphics[width=0.8\textwidth]{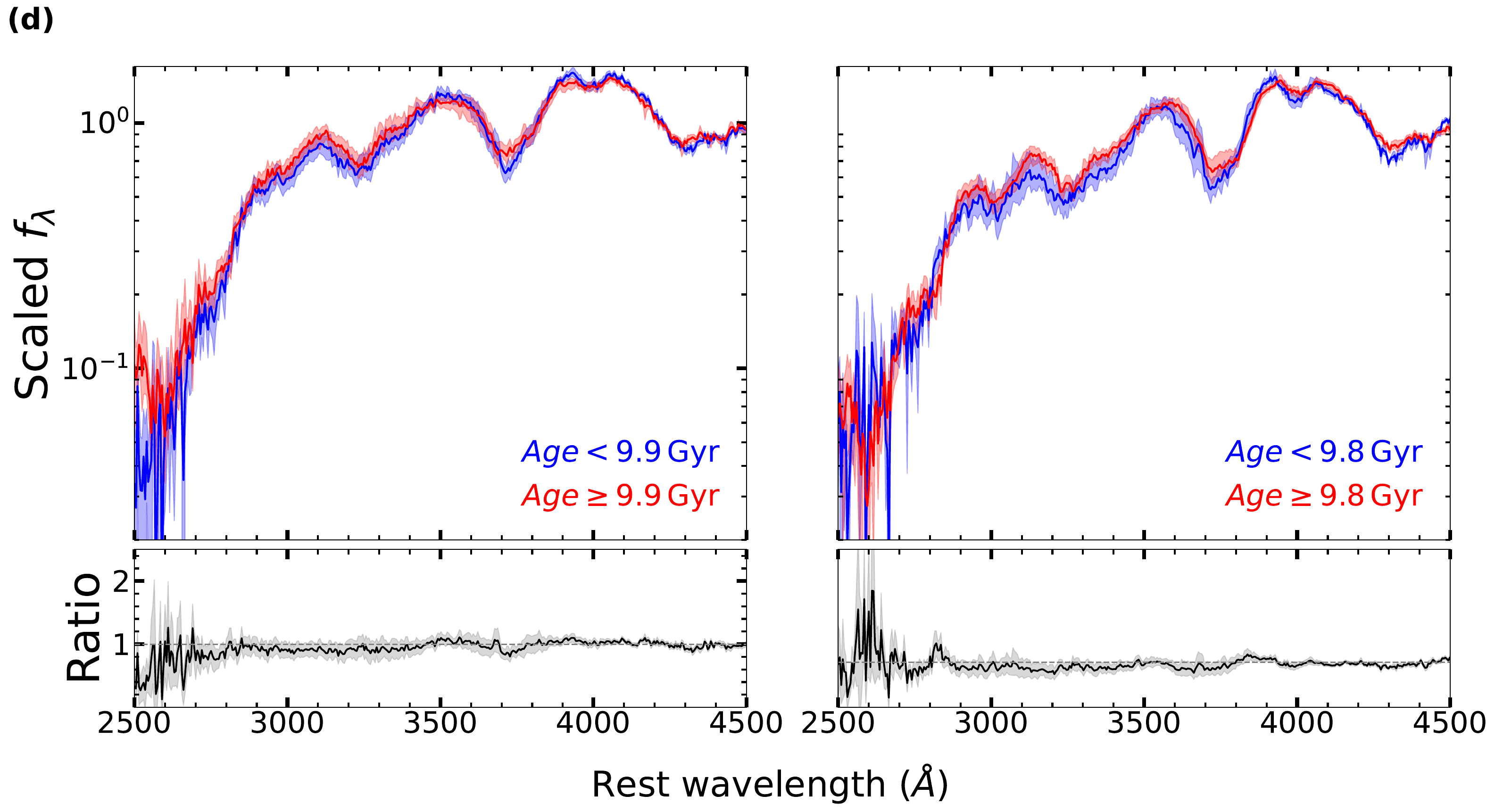}
  \includegraphics[width=0.8\textwidth]{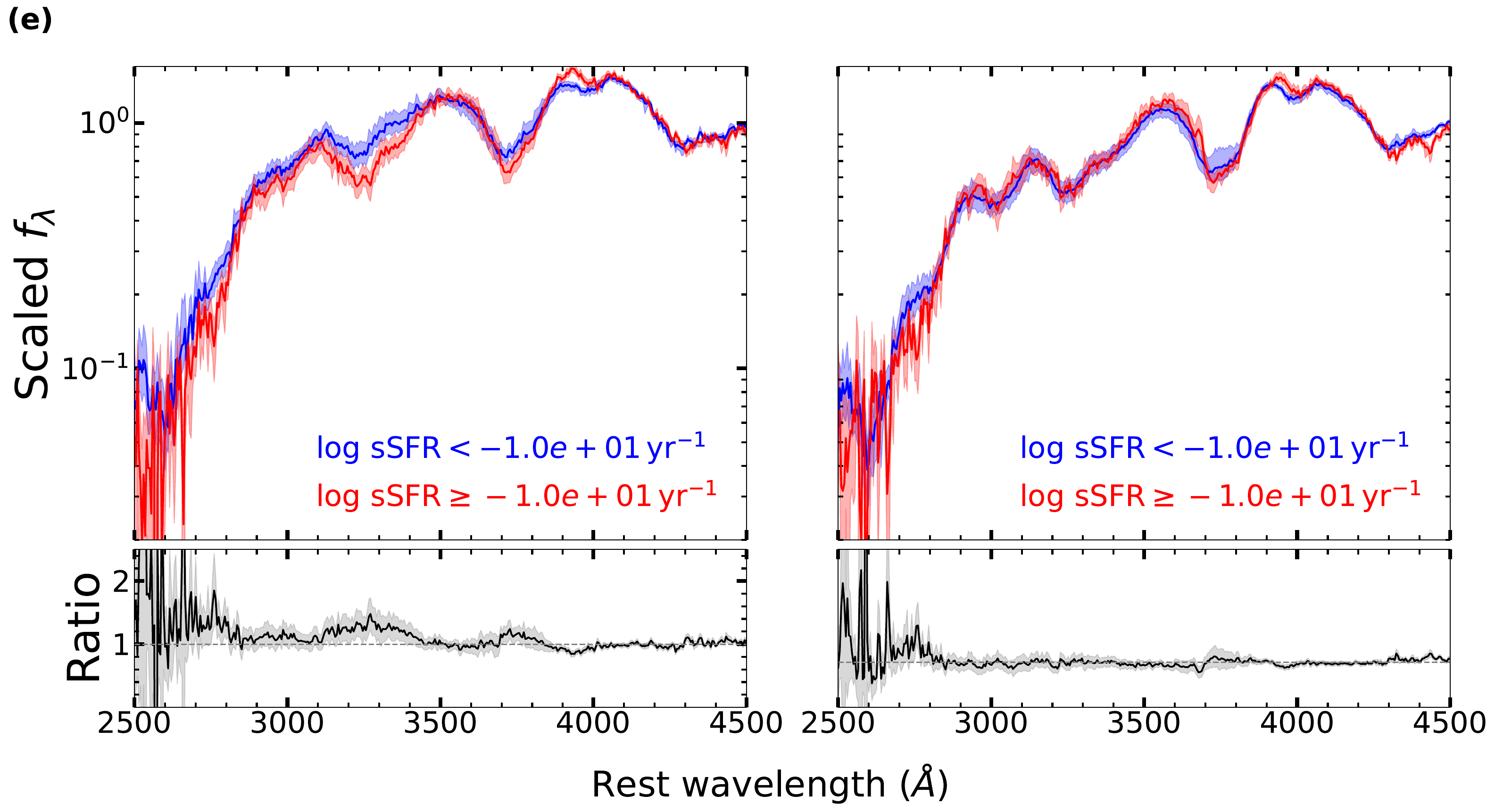}
  \caption{(d) Same as (a) but with respect to host-galaxy stellar age. (e) Same as (a) but with respect to host-galaxy sSFR.}
\end{figure*}

\begin{figure*}[htbp]\ContinuedFloat
  \centering
    \includegraphics[width=0.8\textwidth]{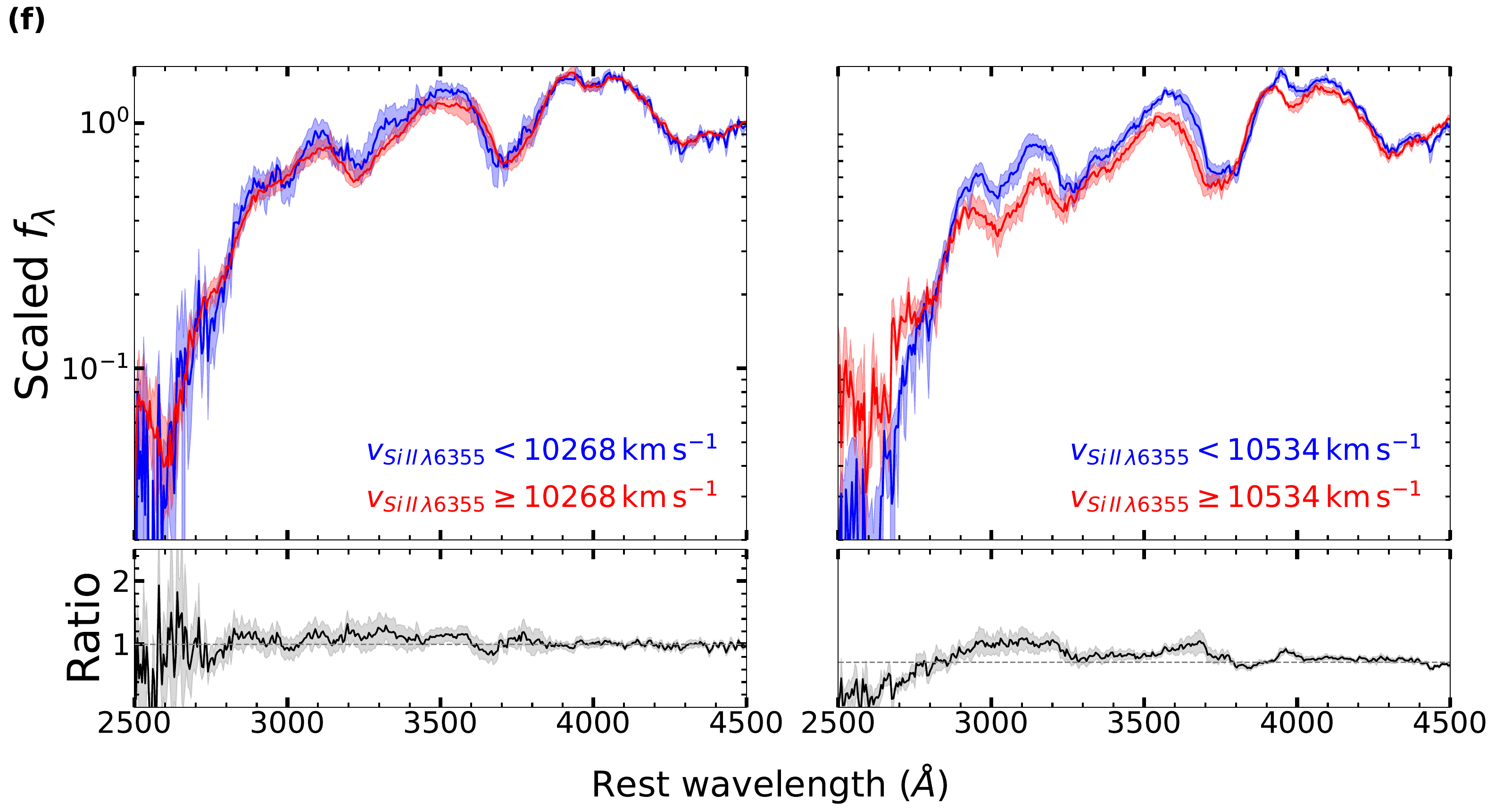}
  \includegraphics[width=0.8\textwidth]{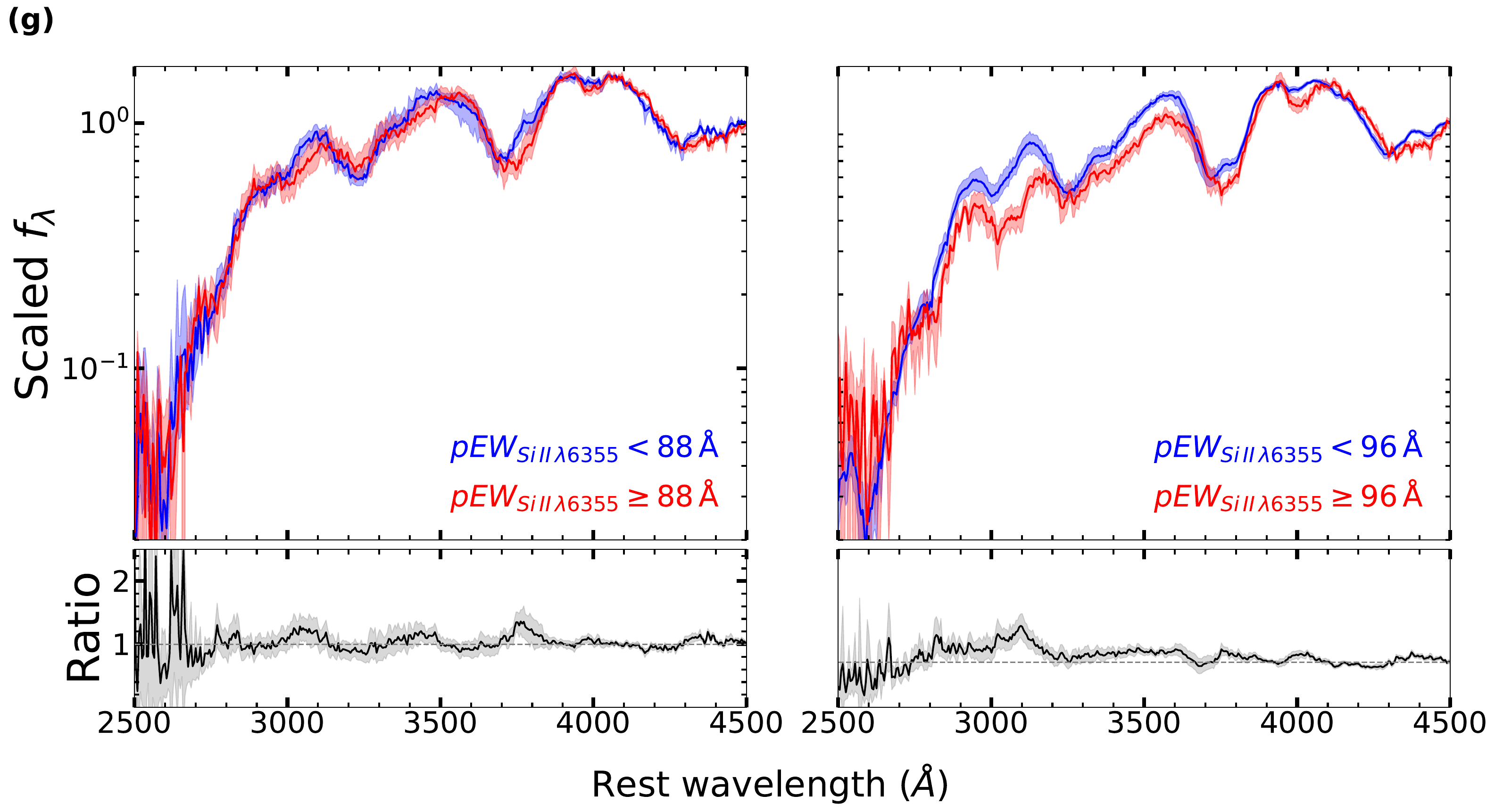}
  \caption{(f) Same as (a) but with respect to the photospheric velocity of \si. (g) Same as (a) but with respect to \si\ pEW.}
\end{figure*}

\subsection{UV flux ratio}
\label{section:flux-ratio}

Here, we quantify the trends observed in Section~\ref{sec:mean} with the flux ratios of UV spectra. The concept of spectral flux ratio is similar to the broad-band color, but with a much narrower spectral region. Following \citet{foley2016} and \citet{pan2020}, we define the flux ratio $f_{2535}$ as the median flux in 2450-2620\,\AA\ (in the MUV) divided by the median flux in 4000-4500\,\AA. Likewise, $f_{3025}$ is defined as the median flux in 2900-3150\,\AA\ (in the NUV) divided by the median flux in the same 4000-4500\,\AA. The uncertainties in the flux ratios are propagated from the uncertainties in flux, wavelength, and host-galaxy reddening using Monte Carlo realizations, as described in detail in \citet{bhattacharjee2025}. The results are presented in Figure~\ref{fig:ratio}, where the colored shades represent the 1-$\sigma$ confidence region of the best-fitted slopes. We only show the confidence region in the figure for cases with possible trends. The analysis follows the same sample selection and classification as in Section~\ref{sec:mean}, with peculiar SNe~Ia excluded.

\begin{figure*}[htbp]
  \centering
  \includegraphics[width=0.79\textwidth]{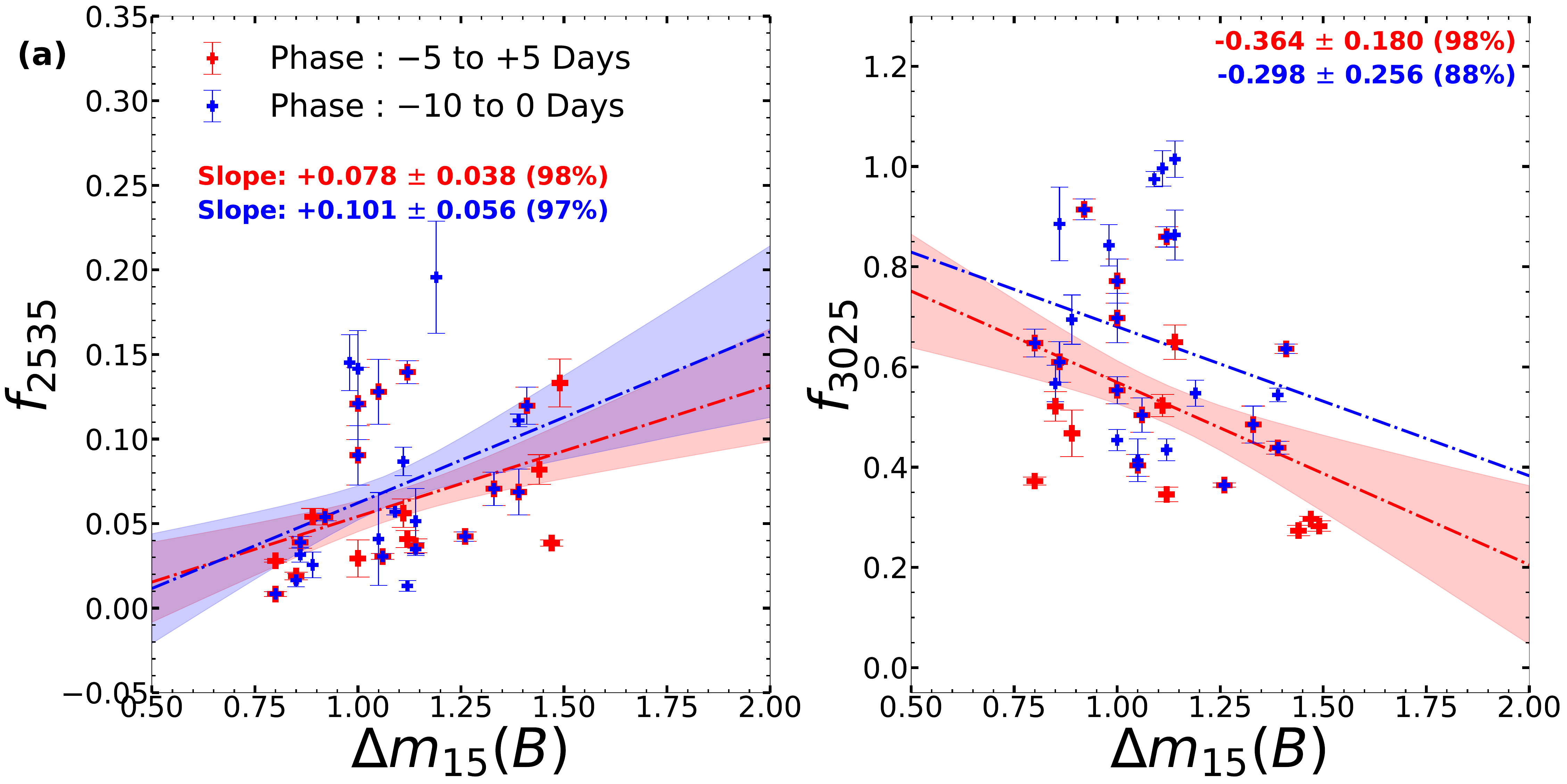}
  \includegraphics[width=0.79\textwidth]{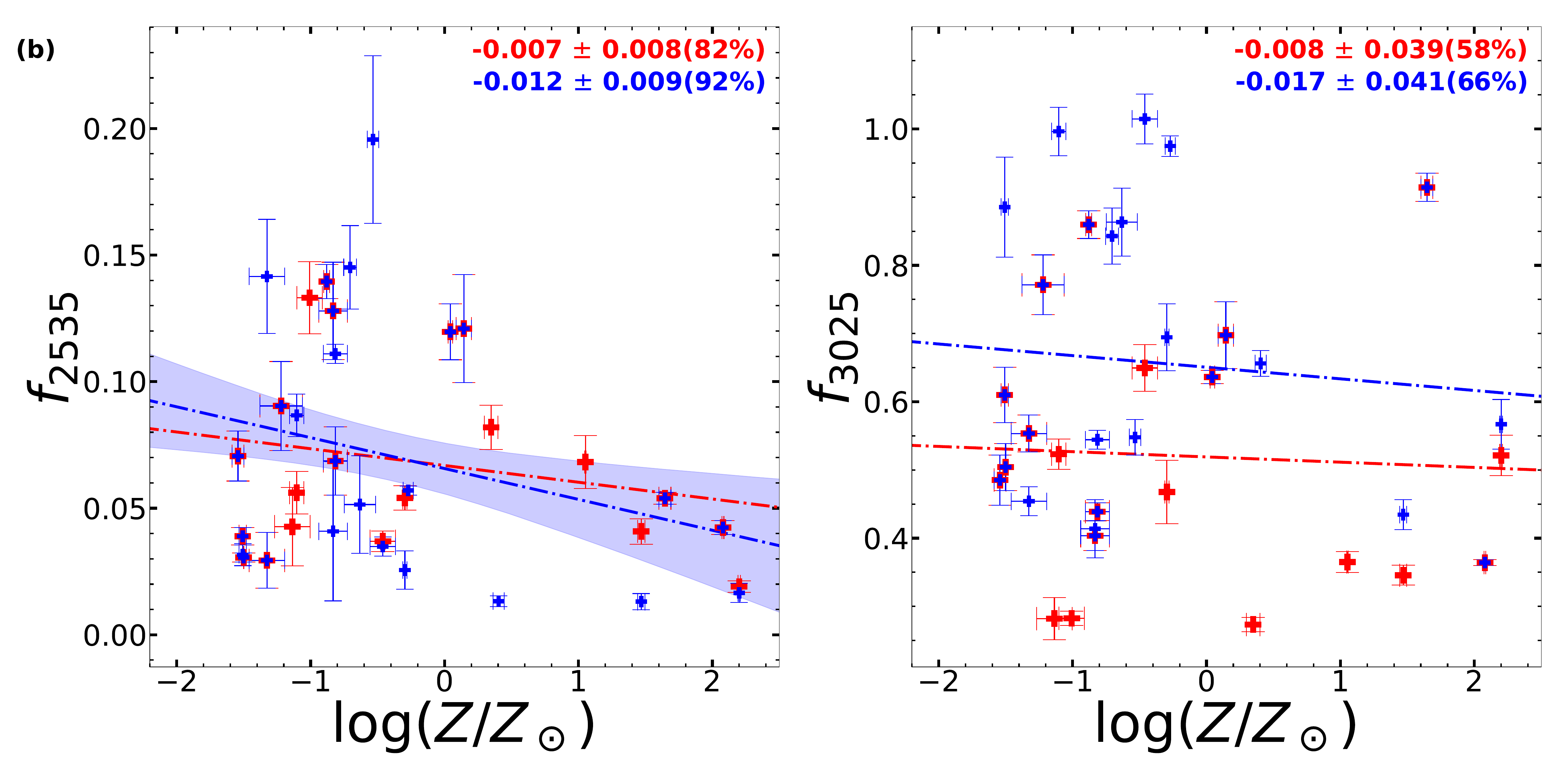}
  \includegraphics[width=0.79\textwidth]{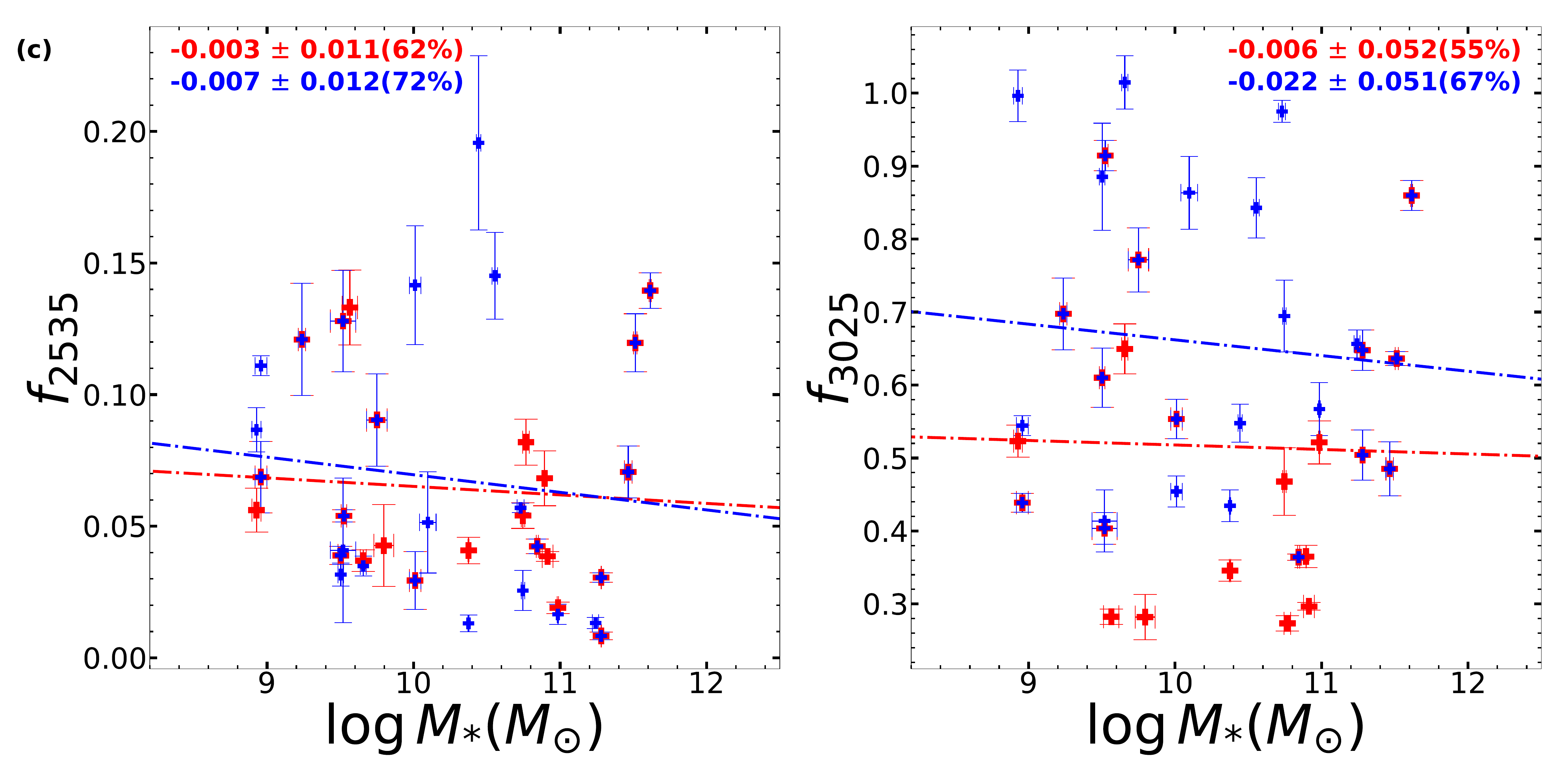}
  \caption{(a) \textit{Left}: The flux ratio $f_{2535}$ as a function of $\Delta m_{15}(B)$. SNe~Ia with pre-maximum observations (i.e., $-10$\,d $\leq$ phase $\leq$ $ 0$\,d) are shown in blue, while near-peak (i.e., $-5$\,d $\leq$ phase $\leq$ $ +5$\,d) SNe are shown in red. Linear fits to these groups are represented by blue and red dashed-dotted lines, respectively, with the corresponding slopes displayed in the panel. The values in brackets indicate the probability of slopes of the same sign as the best-fit slope. The colored shades represent the 1-$\sigma$ confidence region of the best-fitted slopes. We only show the confidence region for those with possible trends. \textit{Right}: Same as left but for the flux ratio $f_{3025}$. (b) Same as (a) but with respect to host-galaxy $Z$. (c) Same as (a) but with respect to host-galaxy $M_{*}$.}
  \label{fig:ratio}
\end{figure*}

\begin{figure*}[htbp]\ContinuedFloat
  \centering
    \includegraphics[width=0.8\textwidth]{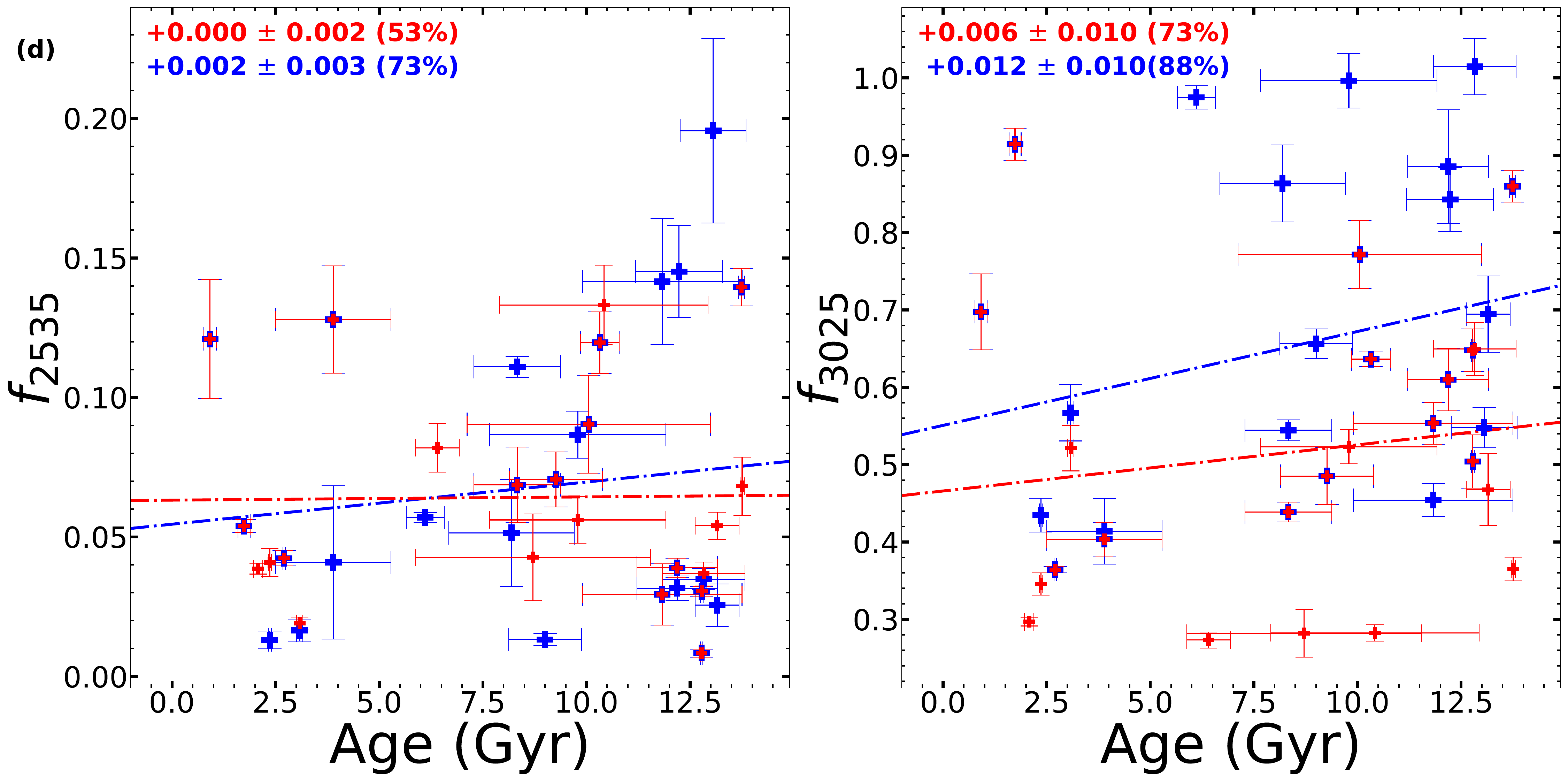}
  \includegraphics[width=0.8\textwidth]{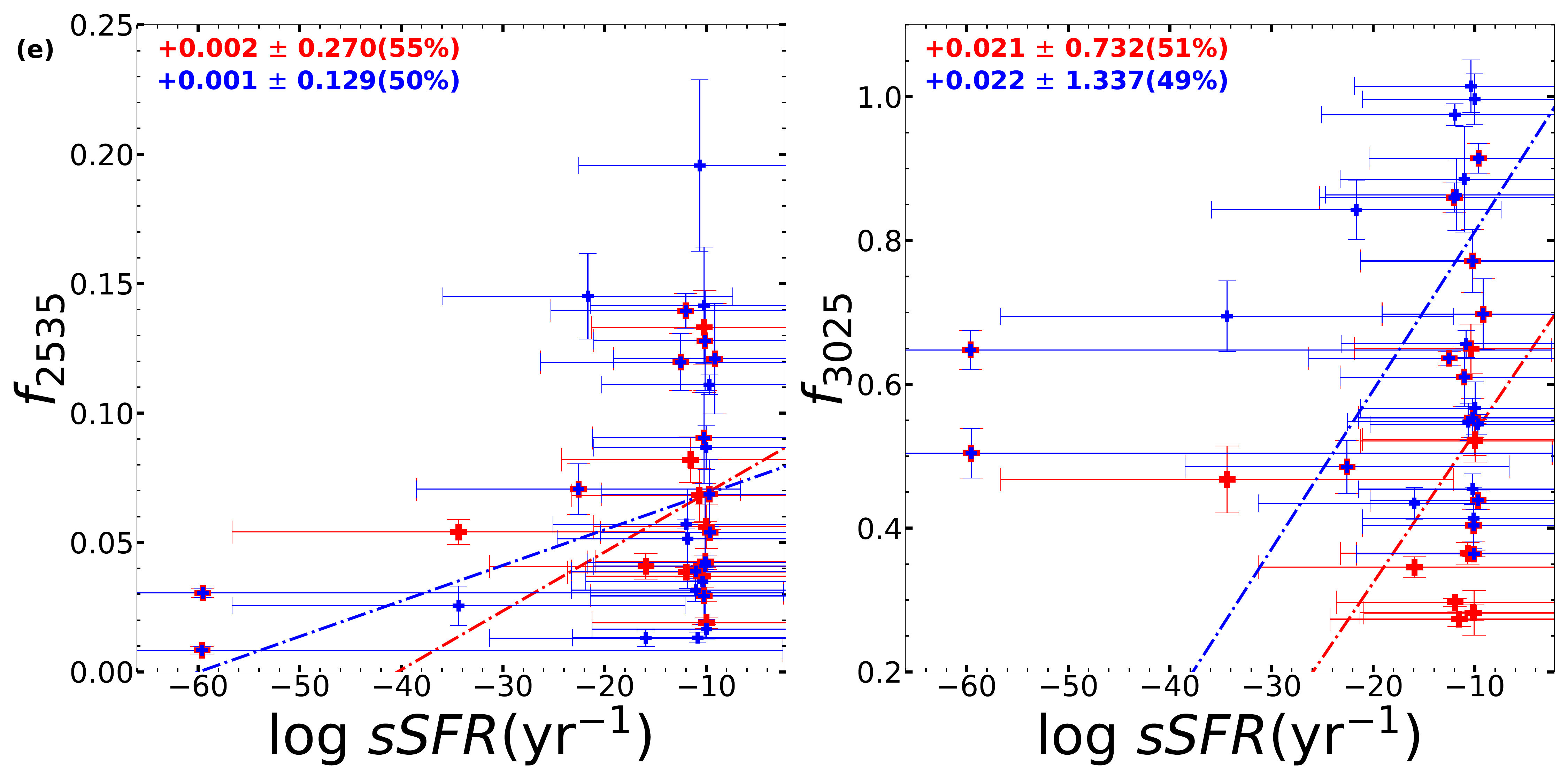}
  \caption{(d) Same as (a) but with respect to host-galaxy stellar age. (e) Same as (a) but with respect to host-galaxy sSFR. }
\end{figure*}

\begin{figure*}[htbp]\ContinuedFloat
  \centering
  \includegraphics[width=0.8\textwidth]{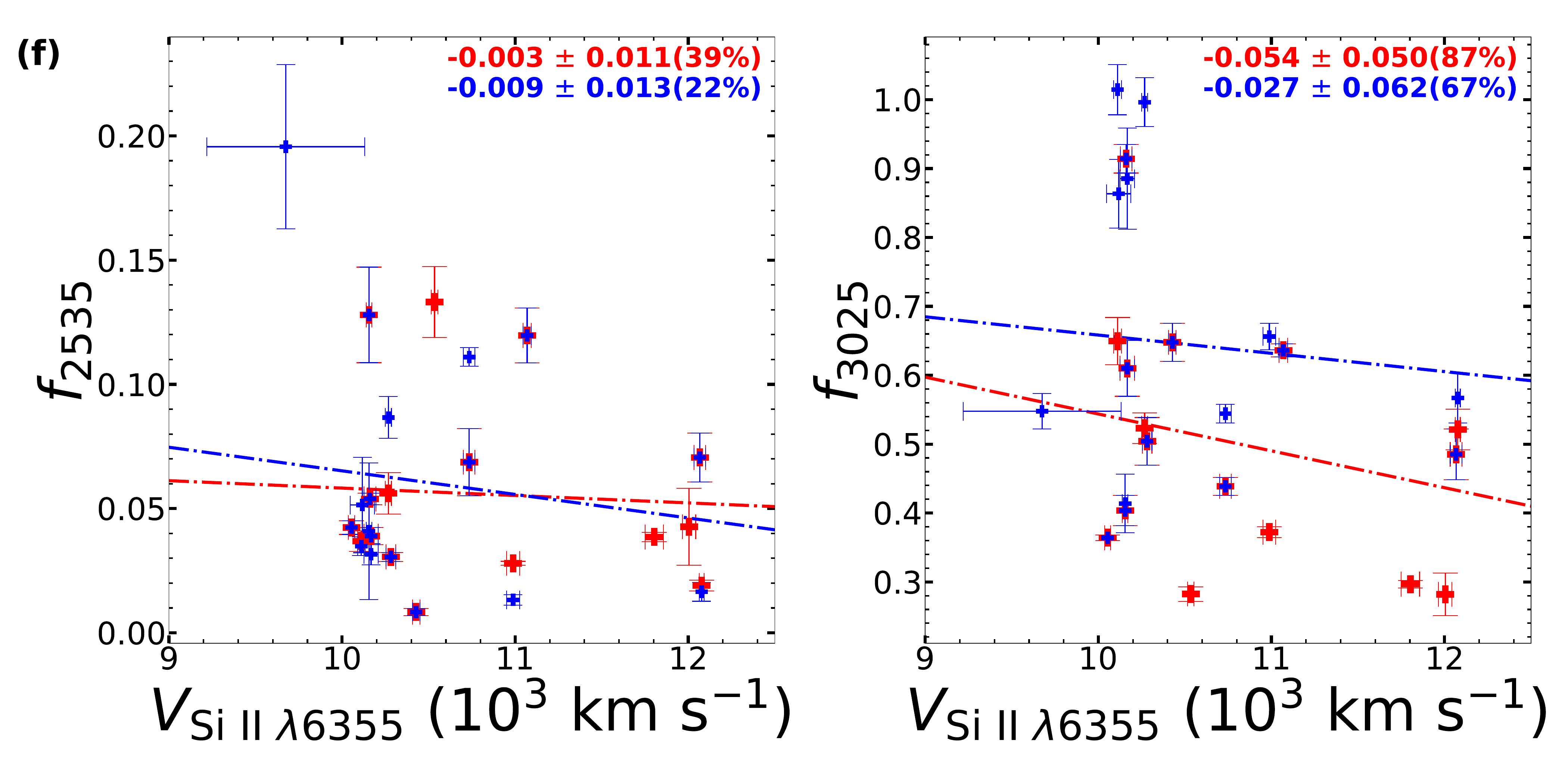}
  \includegraphics[width=0.8\textwidth]{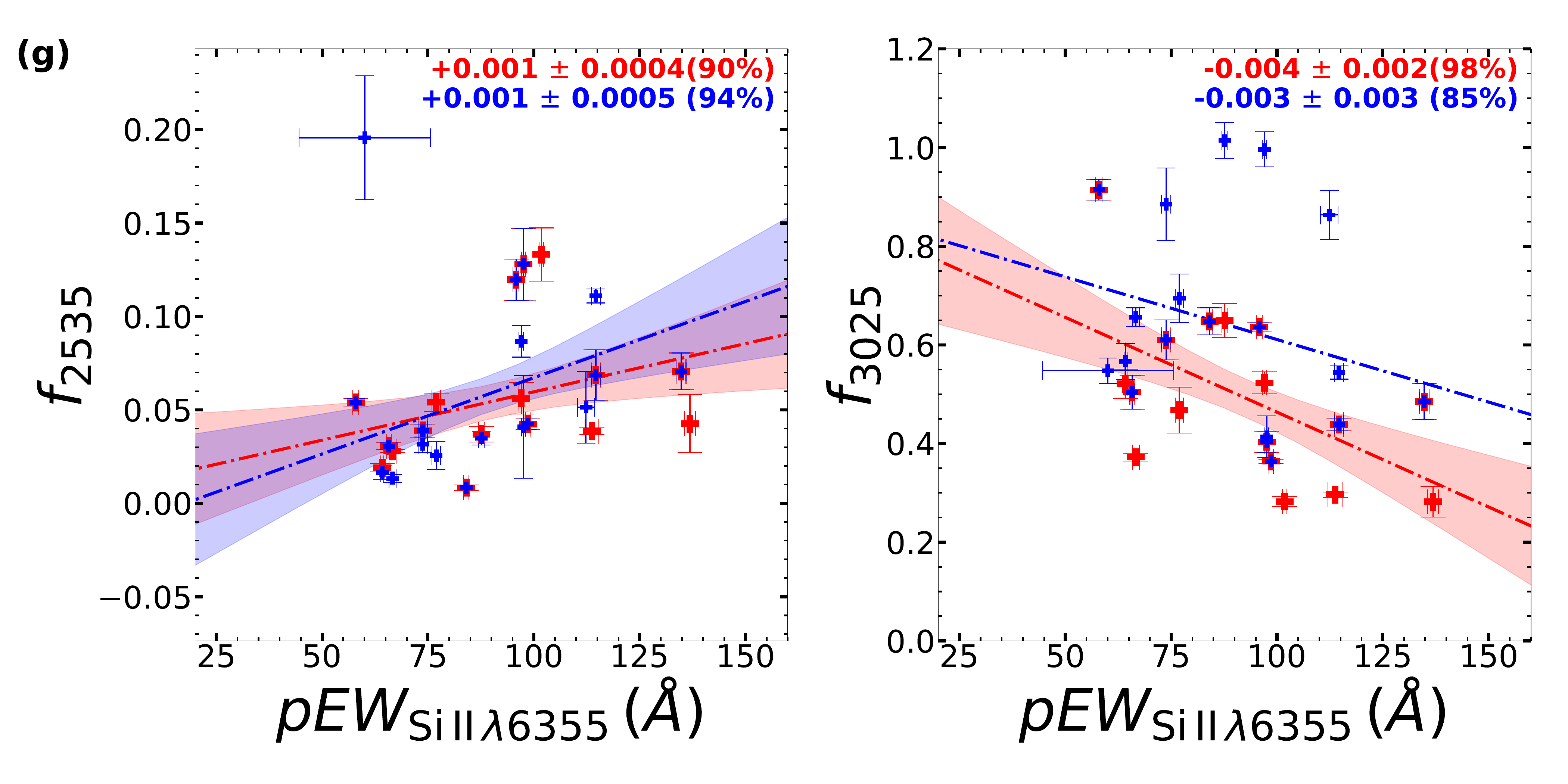}
  \caption{(f) Same as (a) but with respect to the photospheric velocity of \si. (g) Same as (a) but with respect to \si\ pEW.}
\end{figure*}

Panel (a) of Figure~\ref{fig:ratio} shows the results for $\Delta m_{15}(B)$. We find a positive correlation between $f_{2535}$ and $\Delta m_{15}(B)$ for both the pre-maximum and near-peak samples. Using the \texttt{Linmix} package \citep{kelly2007} with 20{,}000 MCMC realizations, we find that $98$\% and $97$\% of the fitted slopes are positive for the pre-maximum and near-peak samples, respectively. In contrast, $f_{3025}$ exhibits an opposite trend in both samples, in the sense that SNe with larger $\Delta m_{15}(B)$ tend to have lower $f_{3025}$. However, this trend is statistically significant only for our near-peak sample, with more than $98$\% of the slopes being negative.

Panels (b) and (c) show the dependence on host-galaxy $Z$ and $M_{*}$, respectively. For $f_{2535}$, we find a possible dependence on $Z$ in the pre-maximum sample, with only $92$\% of the fitted slopes being negative, while no significant trend is seen in the near-peak sample. In contrast, \citet{pan2020} reported a $95$\% correlation based on a smaller sample of $26$ near-peak SNe~Ia. We note that the metallicities presented in this work correspond to stellar metallicities derived from host-galaxy broad-band photometry, while \citet{pan2020} measured gas-phase metallicities from host-galaxy spectra. We also restrict our sample to the same SNe studied in \citet{pan2020}, and still find no significant correlation. This suggests that the discrepancy could arise from differences in the measurement methods. We find no significant trend between the flux ratios and host-galaxy $M_{*}$, consistent with the results of \citet{pan2020}.

Panels (d) and (e) show the results for host-galaxy stellar age and sSFR, respectively. In both cases, we do not observe any significant correlations with the flux ratios. The sSFR is also expected to correlate with galaxy age, in the sense that galaxies with higher sSFR tend to host younger stellar populations than those with lower sSFR \citep[e.g.,][]{Brinchmann2004}. It is therefore not surprising that sSFR and galaxy age show consistent results.

Panel (f) shows the results for the \si\ velocity, where we do not find any significant correlation, consistent with the results of \citet{foley2016}. However, our sample contains relatively few HV SNe~Ia, which are known to differ systematically from NV SNe~Ia in several properties \citep[e.g.,][]{wang2009,Wang2013,Wang2019}. A larger sample including more HV objects may therefore be necessary to fully understand potential trends between the UV flux and ejecta velocity. Panel (g) presents the results for the \si\ pEW. We find a positive correlation with $f_{2535}$ for the pre-maximum sample, with $94\%$ of the best-fit slopes being positive, and a weaker trend with the near-peak sample, with $90\%$ of the slopes being positive. This suggests that SNe with higher UV flux levels near $2500$\,\AA\ tend to exhibit stronger \si\ absorption. In contrast, we observe a statistically significant anti-correlation with $f_{3025}$ for the near-peak sample, with $98\%$ of the slopes being negative. These trends are broadly consistent with those found for $\Delta m_{15}(B)$, which is not surprising given that SNe~Ia with larger $\Delta m_{15}(B)$ tend to have larger \si\ pEW values \citep[e.g.,][]{zhao2021}.

As a sanity check, we also repeat the analysis using an earlier and narrower phase range of $-10$ to $-5$\,days for the pre-maximum sample. The resulting trend is generally consistent with that obtained using the $-10$ to $0$\,day range, although with larger scatter, likely due to the substantially smaller sample size (reduced by a factor of two).

\subsection{SNe~Ia with early-time light-curve excess}\label{excess}

Early-time photometric observations reveal that some SNe~Ia show deviations from the typical power-law rise \citep[e.g.,][]{Marion2016,Jiang2017,Hosseinzadeh2017,Shappee2018,Dimitriadis2019,Li2019,Shappee2019,Ashall2022,Wang2024}. These objects show a flux excess above the expected rise, and have been observed across both normal and peculiar subtypes of SNe~Ia. Several theoretical models have been proposed to explain the excess, including the interaction of SN ejecta with a non-degenerate companion star \citep[e.g.,][]{kasen2010}, interaction with the CSM \citep[e.g.,][]{piro2016}, presence of radioactive $^{56}$Ni close to the surface \citep[e.g.,][]{noebauer2017}, and the double detonation scenario \citep[e.g.,][]{polin2019}.  

\begin{figure*}
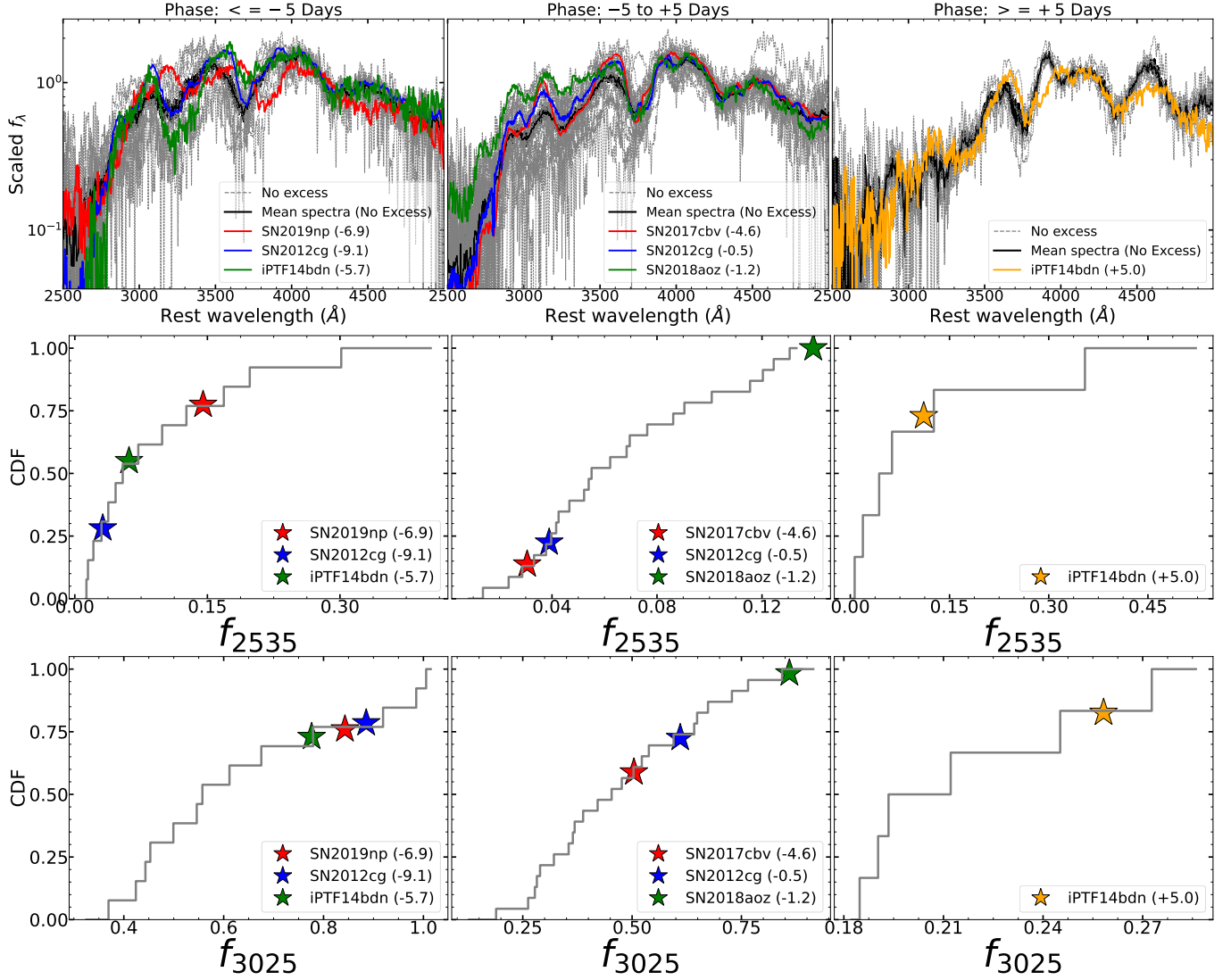

\centering
\includegraphics[width=\textwidth]{compare_bump_double.pdf} 
\includegraphics[width=\textwidth]{cdf_bump.pdf} 
\caption{\textit{Top:} Spectral comparisons between SNe~Ia with early-time excess (in color) and without excess (in grey) with phases earlier than $-5$\,days (left), from $-5$\,days to $+5$\,days (middle), and later than $+5$\,days (right).  All the spectra are normalized with the median flux at 4000--4500\,\AA. The mean spectra of the non-excess comparison sample are shown in black. \textit{Middle:} The cumulative distribution functions (CDFs) of the flux ratio $f_{2535}$ for SNe without excess in the same phase ranges as the top panel, with colored markers indicating the interpolated CDF values for SNe with excesses. \textit{Bottom:} Same as the middle row, but for the flux ratio $f_{3025}$.}\label{bump}
\end{figure*}

Our sample contains $5$ SNe~Ia with reported excess in their early-time light curves: SN~2012cg \citep{Marion2016}, iPTF14bdn \citep{smitka2015}, SN~2017cbv \citep{Hosseinzadeh2017}, SN~2018aoz \citep{Ni2023}, and SN~2019np \citep{sai2022}. Among these, SN~2012cg, SN~2017cbv, SN~2018aoz, and SN~2019np are classified as normal SNe~Ia, while iPTF14bdn belongs to the peculiar 1999aa-like SNe~Ia.  We compare the UV spectra of these objects with those of SNe for which no early-time excess has been reported, although the absence of such a detection may reflect either intrinsic diversity or insufficient early-time coverage. The results are presented in Figure~\ref{bump}, where the top row shows the UV spectral comparisons in three non-overlapping phase ranges, together with the mean spectrum of the non-excess comparison sample in each phase range. The individual non-excess spectra are also shown to illustrate the intrinsic dispersion of the comparison sample. The non-overlapping phase ranges are adopted to ensure that each SN contributes to only one phase range. The middle and lower rows show the cumulative distribution functions (CDFs) of the flux ratios $f_{2535}$ and $f_{3025}$, respectively.

We find that, although the SNe with early-time excess lie within the overall dispersion of the comparison sample, at the early phases ($\leq-5$\,days) all three SNe exhibit higher $f_{3025}$ ratios than a large fraction ($\gtrsim75$\%) of non-excess SNe. A similar trend is seen at maximum light. At later phases ($\geq+5$\,days), however, only one SN with early-time excess is available, and we therefore do not draw any firm conclusion from this phase range. Part of this trend may be related to the decline-rate dependence, since most of these SNe have relatively slow decline rates ($\Delta m_{15}(B)<1.1$ mag) and are therefore expected to show higher $f_{3025}$ values (see Section~\ref{section:flux-ratio}). However, SN 2018aoz is notable for showing a relatively high $f_{3025}$ despite its larger $\Delta m_{15}(B)$ compared to SN~2012cg and SN~2017cbv, suggesting that decline rate alone cannot explain the observed diversity. We note that the nature of the early excess in SN~2018aoz is also different from those of SN~2012cg and SN~2017cbv. While the latter two show blue early excesses, SN~2018aoz exhibits a short-lived red excess during the infant phase \citep{Ni2023}. This suggests that SNe~Ia showing early excesses may not necessarily share the same physical origin and therefore need not show similar UV spectral properties. The larger $\Delta m_{15}(B)$ of SN 2018aoz may partly account for its relatively high $f_{2535}$, given the positive correlation between $f_{2535}$ and decline rate, but it is unable to explain its high $f_{3025}$, for which the overall trend is in the opposite direction. Additional parameters such as host-galaxy properties may also contribute (see Section~\ref{sec:discuss} for a discussion).

We note that this behavior differs from that observed in 03fg-like SNe~Ia, which show a broad UV enhancement across multiple wavelength regions \citep{bhattacharjee2025}. The early-excess SNe in our sample do not exhibit such a global UV flux excess, suggesting that their early photometric excesses may originate from different physical mechanisms, rather than being dominated by strong CSM interaction. Alternatively, any CSM interaction in these objects may be weaker than that inferred for 03fg-like events. The current sample remains too small to investigate the diversity of early-excess properties, such as the morphology, duration, and wavelength dependence of the excess emission. A larger sample is needed to further investigate the physical origins of these events.

\subsection{UV spectra of peculiar SNe~Ia}

SNe~Ia do not comprise a completely homogeneous class, as several subtypes display distinct photometric and/or spectroscopic properties compared to normal SNe~Ia. These subtypes include overluminous events such as SN~1991T-like (hereafter 91T-like; e.g., \citealt{Filippenko1992}), SN~1999aa-like (99aa-like; e.g., \citealt{Garavini2004}), and 03fg-like as well as subluminous events such as SN~1991bg-like (91bg-like; \citealt{Filippenko1992AJ,Leibundgut1993}), SN~2002cx-like \citep[e.g.,][]{Li2003,Phillips2007}, and SN~2002es-like SNe \citep[e.g.,][]{Ganeshalingam2012}. Unlike normal SNe~Ia, many of these peculiar subtypes do not follow the Phillips relation \citep{phillips1993,phillips1999}, and therefore are unsuitable for cosmological usage. Characterizing their UV spectra is key for distinguishing peculiar events whose inclusion without proper treatment could complicate cosmological distance measurements, particularly at high redshift, where the rest-frame UV data play a critical role.

\begin{figure*}
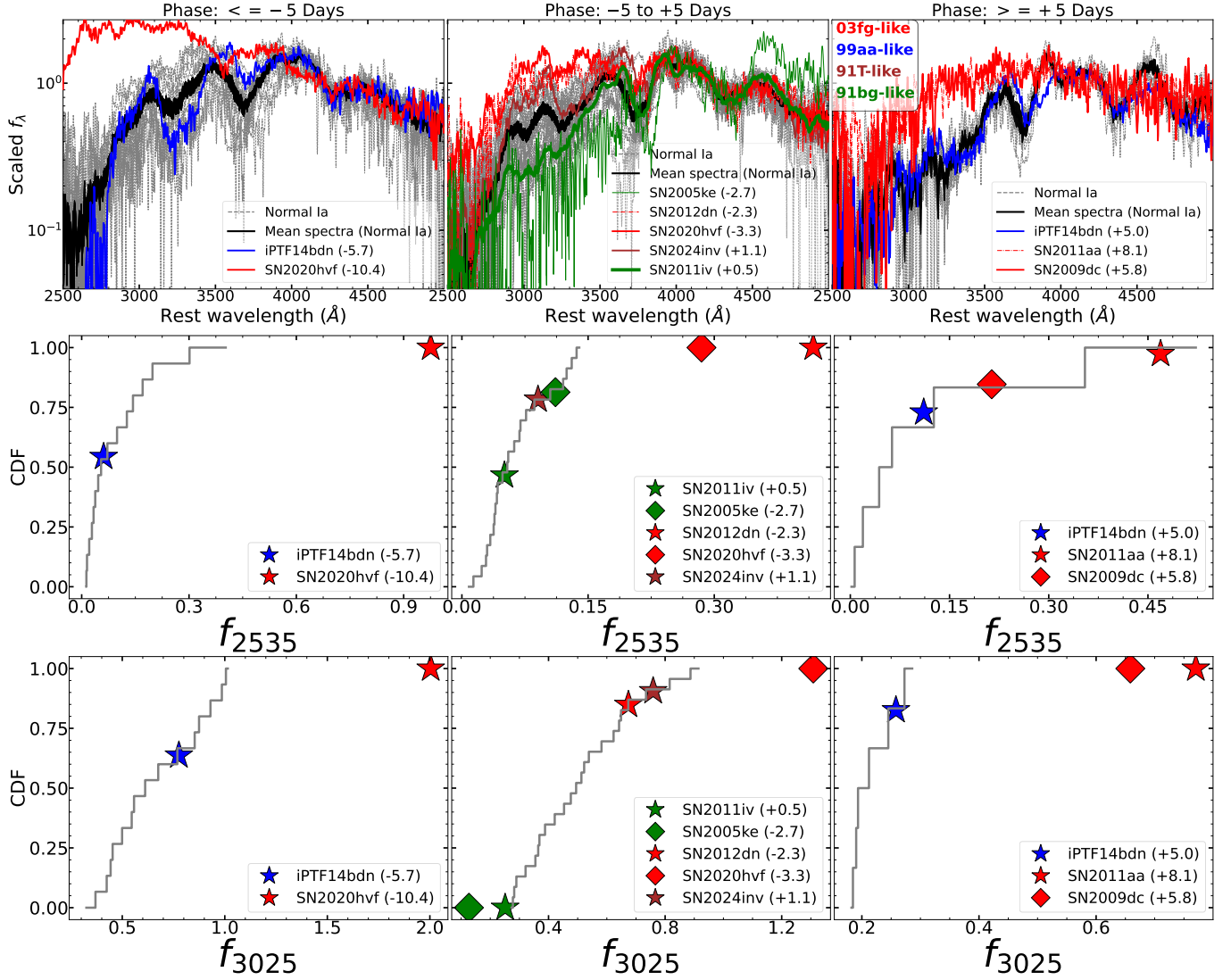

\centering
\includegraphics[width=\textwidth]{compare_subtype.pdf} 
\includegraphics[width=\textwidth]{cdf_subtype.pdf} 
\caption{\textit{Top:} Spectral comparisons between 03fg-like (red), 99aa-like (blue), 91T-like (brown), 91bg-like (green), and normal SNe~Ia (gray) with phases earlier than $-5$\,days (left), from $-5$\,days to $+5$\,days (middle), and later than $+5$\,days (right). All the spectra are normalized with the median flux over 4000--4500\,\AA. The mean spectra of the normal SNe~Ia comparison sample are shown in black. \textit{Middle:} Cumulative distribution functions (CDFs) of the flux ratio $f_{2535}$ for normal SNe in the same phase ranges as in the top row, with colored markers indicating the interpolated CDF values for the peculiar SNe. \textit{Bottom:} Same as the middle row, but for the flux ratio $f_{3025}$.}\label{subtype}
\end{figure*}

Our sample includes two 91bg-like SNe (SN~2005ke and SN~2011iv), one 99aa-like SN (iPTF14bdn), and one 91T-like SN (SN~2024inv). In addition, we incorporate three 03fg-like SNe (SN~2020hvf, SN~2009dc, and SN~2012dn), which have been extensively studied in \citet{bhattacharjee2025}. For these 03fg-like events, we do not perform detailed spectral analyses. They are included solely to provide a more complete comparison across different subtypes. The results are presented in Figure~\ref{subtype}. The top row shows the UV spectral comparisons for each subtype across the same phase ranges adopted in Section~\ref{excess}, while the middle and bottom rows display the CDFs of the flux ratios $f_{2535}$ and $f_{3025}$, respectively.

As previously reported by \citet{bhattacharjee2025}, the 03fg-like events, which represent one of the brightest classes of thermonuclear explosions, tend to exhibit significantly higher UV flux levels compared to normal SNe~Ia. For other subclasses, in general, we find that the brighter and hotter subclasses tend to show higher UV flux levels than their cooler and fainter counterparts in the NUV region. However, this trend is not always observed at shorter wavelengths, as probed by $f_{2535}$, again suggesting that this region could be governed by a more complex combination of physical parameters. We note that this trend should be interpreted primarily in the context of the subclasses examined in this comparison, such as 91T-like, 99aa-like, and 91bg-like SNe~Ia. Other peculiar subclasses may not follow the same trend. For example, previous studies have suggested that 02es-like and Type Iax SNe can exhibit relatively blue UV colors, at least at early phases \citep{Milne2010,hoogendam2024}, despite their lower luminosities, although UV spectroscopic observations for these subclasses remain very limited.

\section{Discussions}\label{sec:discuss}

\subsection{Relation between UV spectra and decline rate}

Although spectroscopically normal SNe~Ia are well studied, uncertainties persist regarding their progenitor systems and explosion mechanisms. UV spectra provide a sensitive probe of the outer ejecta, whose composition, ionization state, and density structure are influenced by both the progenitor system and the explosion physics \citep[e.g.,][]{sauer2008,walker2012,brown2015,lentz2000,derKacy2020}. Our results suggest that the UV spectral properties of SNe~Ia are likely shaped by multiple physical mechanisms, and that their relative importance varies with wavelength. This wavelength dependence may explain why different UV regions show distinct empirical trends.

Previous studies have shown that the UV flux level around $3000$\,\AA, represented by the flux ratio $f_{3025}$, appears to be closely linked to the SN~Ia decline rate \citep[e.g.,][]{foley2016,pan2020}, in the sense that faster-declining SNe~Ia tend to show lower flux levels in this region. This is supported by radiative-transfer models showing that the NUV colors of SNe~Ia can vary with bolometric luminosity \citep{walker2012,derKacy2020}. In particular, higher-luminosity models have higher temperatures, shifting the dominant species from singly to doubly ionized ions and modifying the UV line opacity. At shorter wavelengths, the far-UV and MUV fluxes are expected to be more strongly affected by iron-group line blanketing. Models have shown that higher progenitor metallicity can suppress the UV flux by enhancing the opacity from metal lines \citep[e.g.,][]{lentz2000}, while reverse fluorescence can redistribute photons from optical to UV wavelengths and may become more efficient when the IGE abundance in the outer ejecta is enhanced \citep{sauer2008}. Observationally, \citet{pan2020} reported a $\sim2\sigma$ correlation between $f_{2535}$ and host-galaxy gas-phase metallicity, arguing that this wavelength region may contain information about progenitor metallicity.

In this work, we revisit these relations with a larger sample. We find that $f_{3025}$ still exhibits a strong negative correlation with $\Delta m_{15}(B)$ near maximum light, consistent with previous results. The observed anti-correlation between $f_{3025}$ and \si\ pEW further supports this interpretation, given the correlation between \si\ pEW and light-curve width \citep[e.g.,][]{zhao2021}. In contrast, $f_{2535}$ shows a strong positive correlation with $\Delta m_{15}(B)$ in both the pre-maximum and near-peak samples, opposite to the trend found for $f_{3025}$. Furthermore, we find no clear trend between $f_{2535}$ and host-galaxy stellar metallicity in our near-peak sample, while the pre-maximum sample shows only a weak dependence. However, we caution that the metallicity measurements in this work may not be robust tracers of the progenitor metallicity, since they are based on global host-galaxy properties.

The opposite behaviour of $f_{2535}$ and $f_{3025}$ with the decline rate suggests a wavelength-dependent behaviour in the UV. Both wavelength regions are affected by the coupled effects of temperature, ionization balance, iron-group line opacity, and reverse-fluorescence redistribution. Around $3000$ \AA, the higher flux levels in lower-$\Delta m_{15}(B)$ SNe may reflect the combined effects of higher temperatures, changes in ionization balance, and more efficient reverse-fluorescence redistribution from redder wavelengths into the UV. At shorter wavelengths, the MUV flux around $2500$ \AA\ may be more strongly affected by the high density of iron-group lines. Since lower-$\Delta m_{15}(B)$ SNe generally synthesize larger amounts of $^{56}$Ni and other IGEs, they may have stronger UV opacity if more of them are mixed into the outer ejecta. Although reverse fluorescence may also redistribute flux back into this wavelength region, the observed lower $f_{2535}$ values in lower-$\Delta m_{15}(B)$ SNe suggest that the opacity effect is likely a dominant factor here. Detailed radiative-transfer models are needed to determine whether this interpretation can fully explain the observed trend.

\subsection{\textsc{tardis} modelling}

\input{tardis_table}

\begin{figure*} 
\centering
\includegraphics[width=0.9\textwidth]{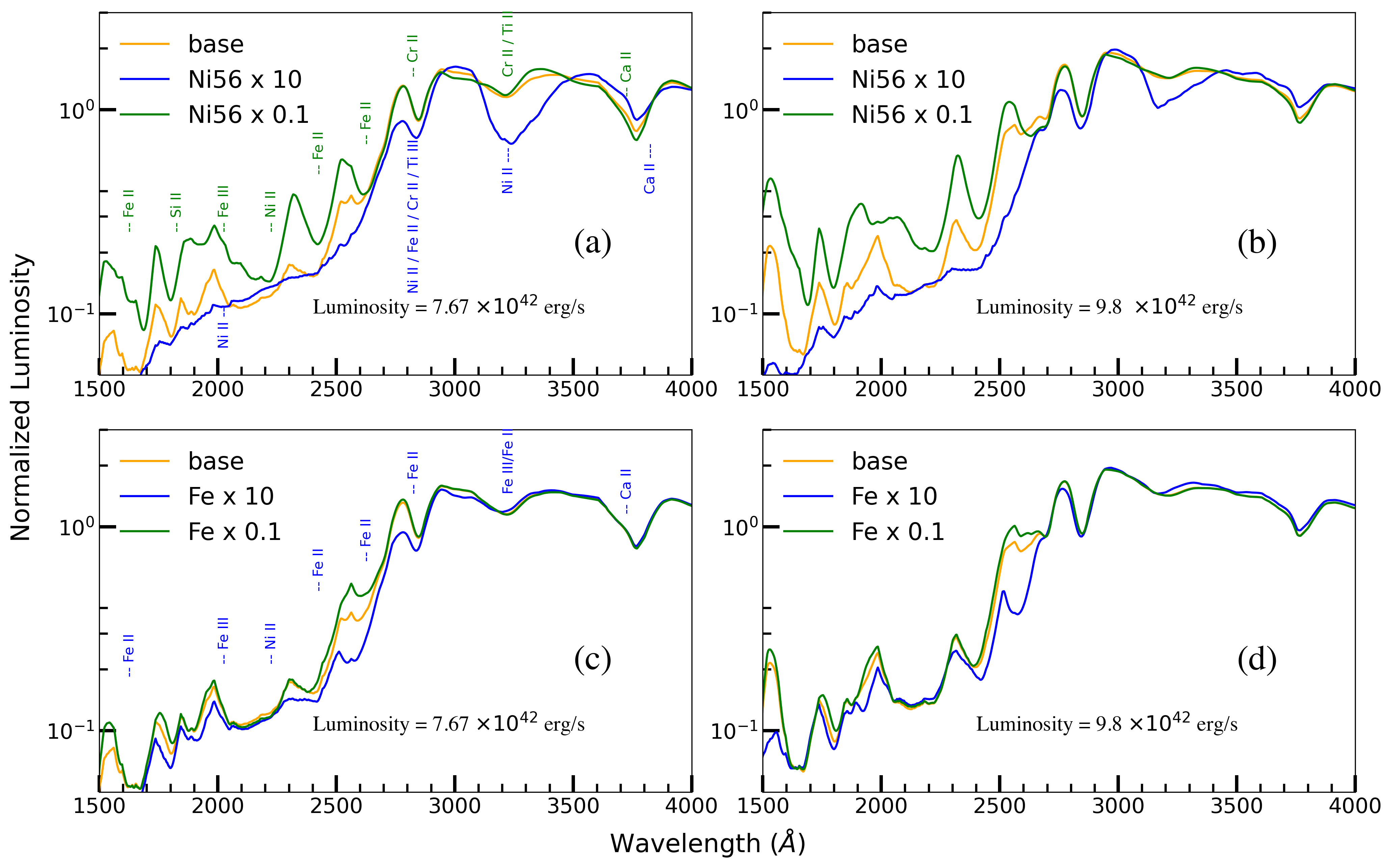} 
\caption{Comparison of UV synthetic spectra generated with \textsc{tardis} for different luminosities and varying $^{56}$Ni and Fe mass fractions. Panels (a) and (b) show models with different $^{56}$Ni mass fractions at lower and higher luminosity, respectively. Panels (c) and (d) show models with different stable Fe mass fractions at lower and higher luminosity, respectively. In panels (a) and (c), we identify the ions contributing to the main spectral features. The ion labels are color-coded according to the corresponding synthetic spectra. The unit of luminosity is in erg~$\rm s^{-1} \AA^{-1}$.} \label{UV-spectra}
\end{figure*}

To understand the origin of the UV spectral properties, we generate spectral models with \textsc{tardis} \citep[][v2025.08.17]{2014MNRAS.440..387K, kerzendorf_2025_16888341}. We adopt a uniform elemental composition above an inner velocity ($v_{inner}$) and a W7-like density profile \citep[similar density profiles have been adopted in previous studies; e.g.,][]{1993A&A...269..423M, 2021MNRAS.503..896D}. Models are computed at $12$\,days after explosion for two luminosities. The inner velocity of the models is placed at $12{,}000$\,$\rm km~s^{-1}$ and the outer velocity at $25{,}000$\,$\rm km~s^{-1}$. We aim to investigate how the UV spectra are affected after varying the outer-layer IGE composition in our synthetic spectra. To do so, we change the mass fractions of radioactive $^{56}$Ni and stable Fe above $v_{\rm inner}$. The mass fractions of all other elements were held fixed, and any change in the IGE abundance is compensated by adjusting the oxygen abundance. The major sources of opacity in the UV are bound-bound transitions and electron scattering \citep{2017hsn..book..769S}. Hence, the models are generated with electron scattering enabled, ionization treated in nebular mode, and excitation in dilute-lte mode. A detailed description of the base models can be found in Table~\ref{tab:tardis_ejecta}. 

Results are illustrated in Figure~\ref{UV-spectra}, where we find that increasing $^{56}$Ni produces a clear suppression of the UV flux levels below $3000$\,\AA, consistent with enhanced UV opacity. We also find that models with higher $^{56}$Ni yield a strong Ni\,{\sc ii} absorption feature at $\sim3200$\,\AA. However, when the $^{56}$Ni mass fraction is reduced to one-tenth of the base value, this feature becomes weaker, and the absorption is instead dominated by Cr\,{\sc ii} and Ti\,{\sc ii}. A similar trend is seen in the observed mean spectra in Section~\ref{sec:mean}, in the sense that the slower-declining SNe~Ia, which are expected to produce larger amounts of $^{56}$Ni, show a stronger absorption feature near $\sim3200$\,\AA\ than their fast-declining counterparts. In contrast, varying the Fe mass fraction has a smaller impact on the synthetic spectra than varying the $^{56}$Ni abundance. Nevertheless, we find a qualitatively similar trend, with higher Fe mass fractions producing lower flux levels below $3000$\,\AA, particularly around $\sim2500$\,\AA.

While these models are computed at fixed luminosity, the trends obtained by varying the $^{56}$Ni abundance are broadly consistent with the wavelength-dependent behavior observed with $\Delta m_{15}(B)$. SNe~Ia with lower $\Delta m_{15}(B)$ generally produce larger amounts of $^{56}$Ni, and our \textsc{tardis} models show that increasing the $^{56}$Ni abundance suppresses the flux at shorter UV wavelengths while enhancing the flux at $\sim3000$\,\AA. This qualitatively matches the observed trends that lower-$\Delta m_{15}(B)$ SNe have lower $f_{2535}$ but higher $f_{3025}$. Therefore, the models suggest that changes in IGE abundance can naturally contribute to the opposite $\Delta m_{15}$ correlations seen in the MUV and NUV regions.

The increase in UV flux may also be related to the reverse fluorescence, where the absorption of redder photons is re-emitted in the UV \citep{2000A&A...363..705M,sauer2008}. \textsc{tardis} allows for such a treatment through the \texttt{macroatom} mode for line interactions \citep[see more in][]{2014MNRAS.440..387K}. We find in our simulations that a significant number of packets undergo reverse fluorescence at their last interaction. However, the line opacity associated with enhanced $^{56}$Ni remains the dominant effect in the MUV in our models.

We caution that fixing the luminosity isolates the effect of $^{56}$Ni abundance, whereas real SNe~Ia with different $\Delta m_{15}(B)$ also differ in luminosity. Therefore, these models should not be interpreted as a direct one-to-one mapping between decline rate and UV spectra. Nevertheless, our comparison between different luminosity models shows that brighter models generally have higher NUV flux levels, while the trend in the MUV is not always monotonic. A broader parameter survey, varying the luminosity, outer-layer density profile, composition, inner boundary velocity, and epoch, is required to better understand the observed UV trends and connect them to the explosion mechanisms. Our \textsc{tardis} models are publicly available on \texttt{Zenodo}.\footnote{\url{https://doi.org/10.5281/zenodo.22646347}}

We emphasize that the abundance variations explored here are intended to investigate the UV diversity among normal SNe~Ia. Peculiar subclasses may involve different physical mechanisms and are not necessarily expected to follow the same trends. In particular, the much stronger and broader UV enhancement observed in 03fg-like events cannot be reproduced by the abundance variations considered here. In \citet{bhattacharjee2025}, we found that their UV excess could instead be reproduced by adding an additional hot blackbody component to a normal SN~Ia spectrum, consistent with an extra source of emission such as CSM interaction. Further observations and dedicated radiative-transfer modeling of other peculiar subclasses will be needed to determine whether their UV properties arise from similar or distinct mechanisms.

\subsection{The host-galaxy dependence}

Although our results do not show a primary dependence of the UV flux on host-galaxy metallicity, this does not necessarily rule out a metallicity effect. The empirical relation between SN~Ia decline rate and host-galaxy properties is not a one-to-one mapping. Metal-rich environments can host both slow- and fast-declining SNe~Ia, while metal-poor environments may preferentially host slower-declining events \citep[e.g.,][]{pan2014}. Therefore, a dependence on decline rate can remain strong even if the trend with host-galaxy metallicity appears weak. In addition, the global host-galaxy metallicity used in this work is not a perfect tracer of the progenitor metallicity. A secondary metallicity effect may therefore still be present, but it may only become visible after controlling for the dominant dependence on light-curve shape.

A useful example is provided by the `twin' SNe~2011fe and 2011by, which are remarkably similar in their optical spectra and light-curve properties but show substantial differences in the UV \citep{foley2013}. \citet{foley2013} argued that the UV difference between these two events could be explained by a difference in progenitor metallicity. This suggests that even among SNe~Ia with very similar optical properties, the UV spectra may still preserve information about progenitor composition.

\begin{figure}
\centering
\includegraphics[width=0.48\textwidth]{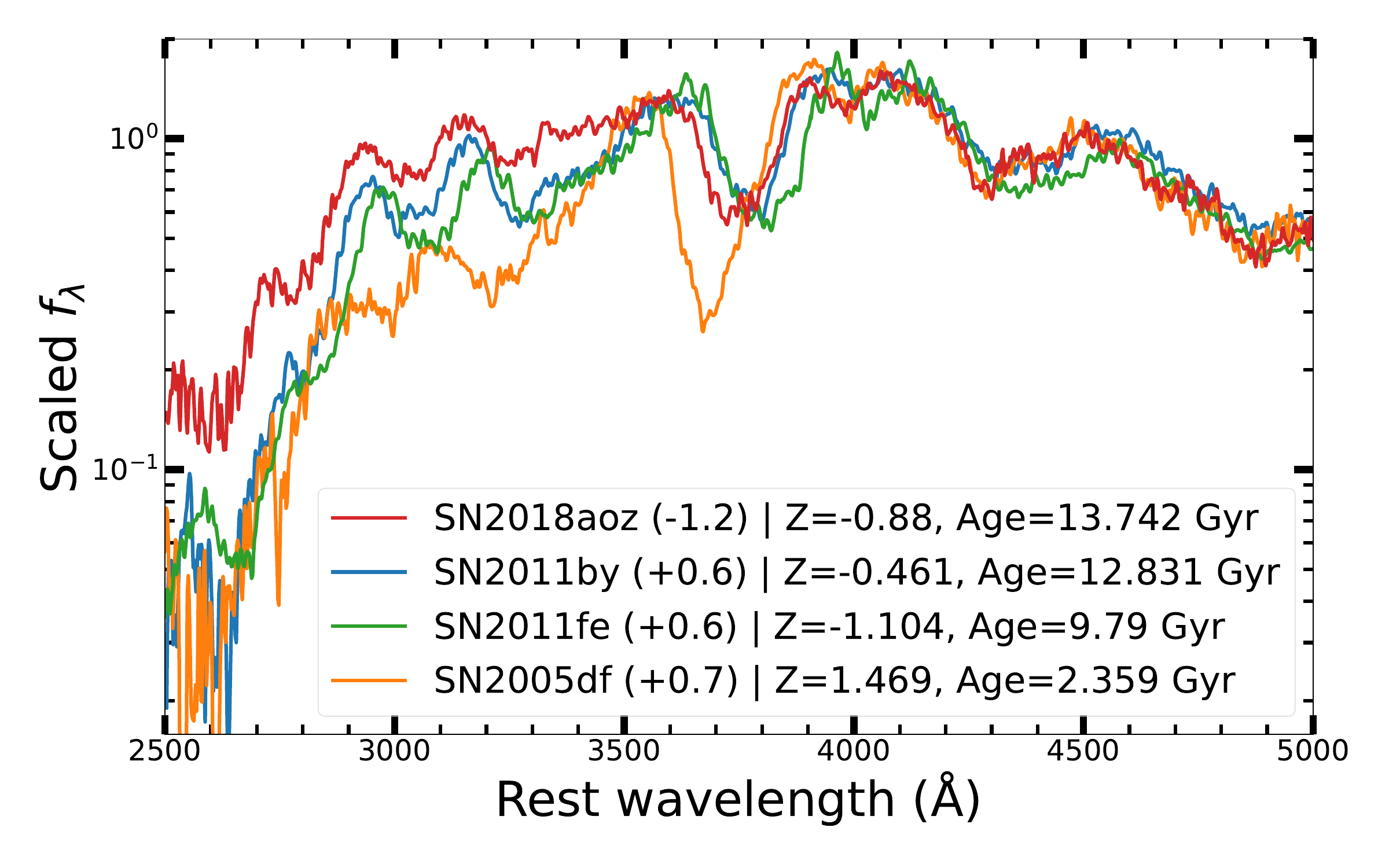} 
\caption{Spectral comparisons between SN~2018aoz, SN~2011by, SN~2011fe, and SN~2005df at similar phases. All these SNe have nearly the same $\Delta m_{15}(B)$. All the spectra are normalized with the median flux over 4000--4500\,\AA.}\label{twin}
\end{figure}

Motivated by this, Figure~\ref{twin} compares the UV spectra of 4 SNe~Ia in our sample having the same $\Delta m_{15}(B)$ ($\approx1.1$) and observed at similar phases. The purpose of this comparison is to examine whether the remaining UV diversity shows any dependence on host-galaxy properties after fixing $\Delta m_{15}(B)$. In this small comparison sample, the UV flux level appears to depend on the host-galaxy properties, with SNe in older and more metal-poor environments exhibiting higher UV flux levels than those in younger and more metal-rich environments. Thus, while host-galaxy properties may play a role, their effect may be secondary to the dominant dependence on light-curve decline rate. Nevertheless, this interpretation should be treated with caution. The sample size is small, and the degeneracy between host-galaxy metallicity and stellar age could also prevent us from identifying the primary driver. Moreover, the trend with host properties could reflect a mixture of multiple progenitor channels, rather than a continuous dependence on the host environment within a single channel. Larger samples with precise host-galaxy measurements will be needed to determine whether the residual UV diversity at a fixed decline rate is primarily driven by metallicity, age, or progenitor-channel differences.

\section{Conclusions}\label{sec:conclude}
In this work, we analyze the UV spectra of a large sample of SNe~Ia from the {\it Swift} UVOT grism observations. This sample is larger than those used in previous works and includes more pre-maximum UV spectra. We investigate relationships between UV spectral properties and various SN and host-galaxy parameters. We also examine the UV spectra of SNe~Ia exhibiting early light-curve excesses, as well as those of peculiar subtypes. Our results are summarized as follows:

\begin{enumerate}
    \item The UV flux levels in the NUV, as traced by the flux ratio $f_{3025}$, tend to decrease with $\Delta m_{15}(B)$, while those in the MUV, as traced by the flux ratio $f_{2535}$, show the opposite trend. This indicates that the relation between UV flux and light-curve decline rate is strongly wavelength dependent. The higher $f_{3025}$ values in lower-$\Delta m_{15}(B)$ SNe may be related to their higher luminosities and temperatures, while the lower $f_{2535}$ values may reflect stronger UV opacity from IGEs.

    \item Our \textsc{tardis} models provide a qualitative explanation for the trends observed with $\Delta m_{15}(B)$. Increasing the $^{56}$Ni abundance suppresses the MUV flux while enhancing the flux in the NUV, broadly consistent with the observed trend with $\Delta m_{15}(B)$, since slow decliners are expected to yield higher amounts of $^{56}$Ni. However, these simplified models are computed at fixed luminosity. A broader parameter survey is needed to fully connect the UV trends to explosion physics.

    \item No statistically significant correlations are found between the UV flux levels and global host-galaxy parameters, such as stellar mass, stellar age, and sSFR. A weak trend is observed only between $f_{2535}$ and host-galaxy stellar metallicity in the pre-maximum sample, indicating that the UV emission may not strongly depend on the global host-galaxy properties. However, comparisons among a small sample of SNe~Ia with similar decline rates reveal a possible residual trend with host-galaxy properties, implying that any host dependence (if present) is likely secondary.

    \item We find a mild difference in the mean UV spectra of lower and higher ejecta-velocity sub-groups, in the sense that the higher-velocity SNe tend to show more blueshifted features and more suppressed flux levels in the NUV, although the flux ratios do not show a significant trend with velocity. The small number of HV SNe~Ia in our sample also prevents us from drawing a firm conclusion. We find that \si\ pEW shows correlations with the flux ratios, in the same direction as the decline rate. These trends are likely related to the dependence of \si\ pEW on the decline rate.

    \item  Spectroscopically normal SNe~Ia exhibiting early-time photometric excesses tend to show enhanced flux mainly in the NUV, rather than a broad enhancement across the UV. This suggests that these excesses may be driven by mechanisms other than strong CSM interaction, or that any CSM interaction is too weak to leave a clear imprint on the UV spectra.

    \item  Different SN~Ia subtypes show distinct UV spectral properties. Among the subclasses examined here, 03fg-like SNe show the strongest UV emission, 99aa- and 91T-like SNe show moderately enhanced UV flux, and 91bg-like SNe exhibit the weakest UV emission. These differences are broadly consistent with their known optical properties. However, we note that the current sample size of peculiar SN~Ia is small and does not include all known subtypes.
    
\end{enumerate}

\section*{Acknowledgements}
YCP is supported by the National Science and Technology Council (NSTC grant 112-2112-M-008-026-MY3). CDK gratefully acknowledges support from the NSF through AST-2432037, the HST Guest Observer Program through HST-SNAP-17070 and HST-GO-17706, and from JWST Archival Research through JWST-AR-6241 and JWST-AR-5441. SB thanks Taiwanese bubble milk tea for the monumental contributions throughout the work.

This research made use of \textsc{tardis}, a community-developed software package for spectral synthesis in supernovae \citep{2014MNRAS.440..387K, kerzendorf_2025_16888341}. The development of \textsc{tardis} received support from GitHub, the Google Summer of Code initiative, and from ESA's Summer of Code in Space program. \textsc{tardis} is a fiscally sponsored project of NumFOCUS. \textsc{tardis} makes extensive use of Astropy. The simulations employed the \texttt{TARDIS} atomic data file named \texttt{kurucz\_cd23\_chianti\_H\_He\_latest.h5}.

\facilities{Swift UVOT}

\software{NumPy \citep{Harris2020},  SciPy \citep{2020NatMe..17..261V}, Matplotlib \citep{2007CSE.....9...90H},
          Pandas \citep{mckinney-proc-scipy-2010}, Linmix \citep{kelly2007}, HostPhot \citep{hostphot2022}, Prospector \citep{johnson2019,johnson2021}, SNooPY \citep{burns2011}, Extinction \citep{Barbary2024_extinction}, Kcorrect \citep{blanton2007}, Tardis \citep{2014MNRAS.440..387K, kerzendorf_2025_16888341}.
          }


\appendix

\section{\texttt{SNooPy} Light Curve fitting}\label{snpy:app}

\begin{figure*}
\centering
\includegraphics[width=0.49\textwidth]{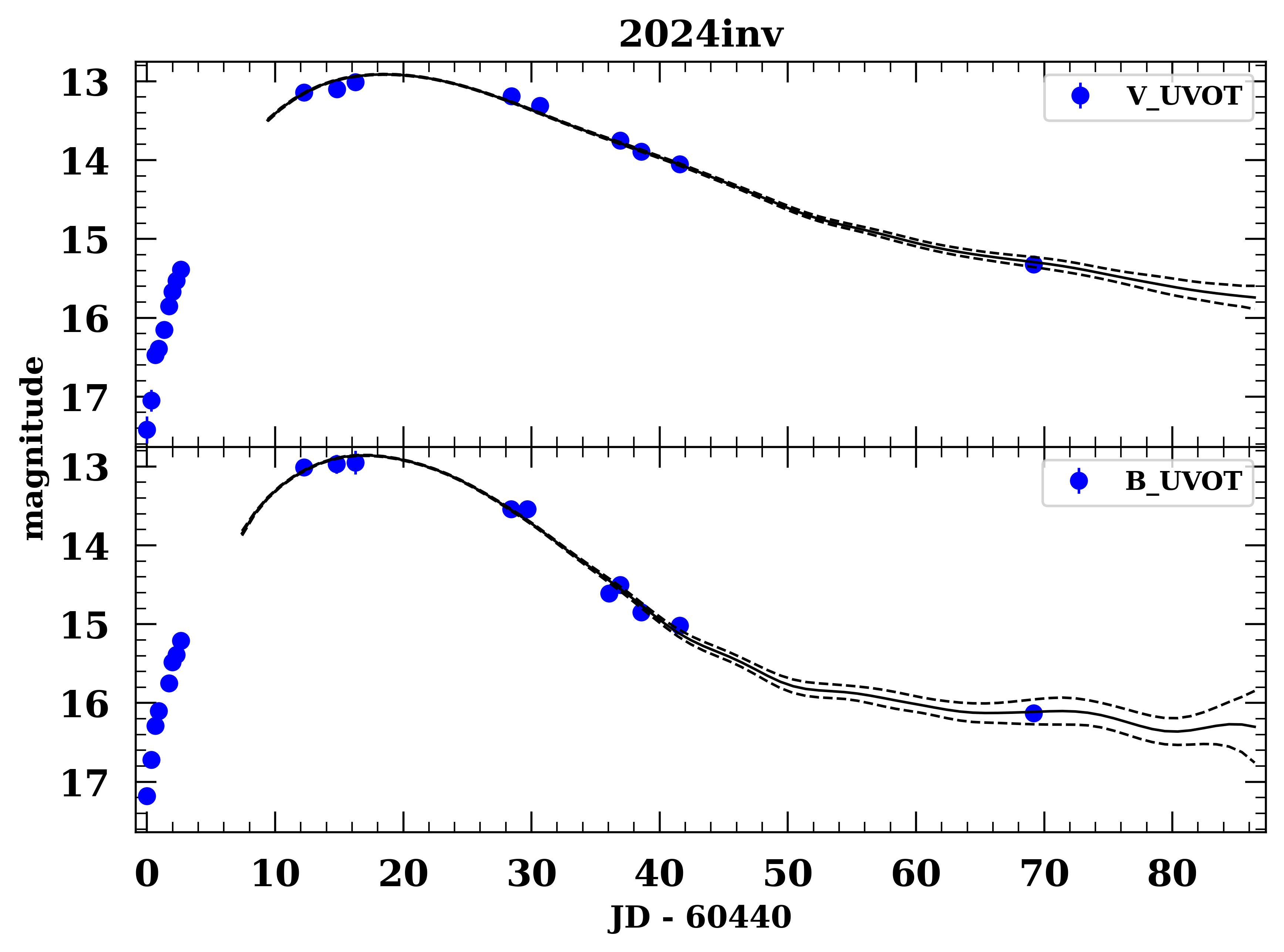}
\includegraphics[width=0.49\textwidth]{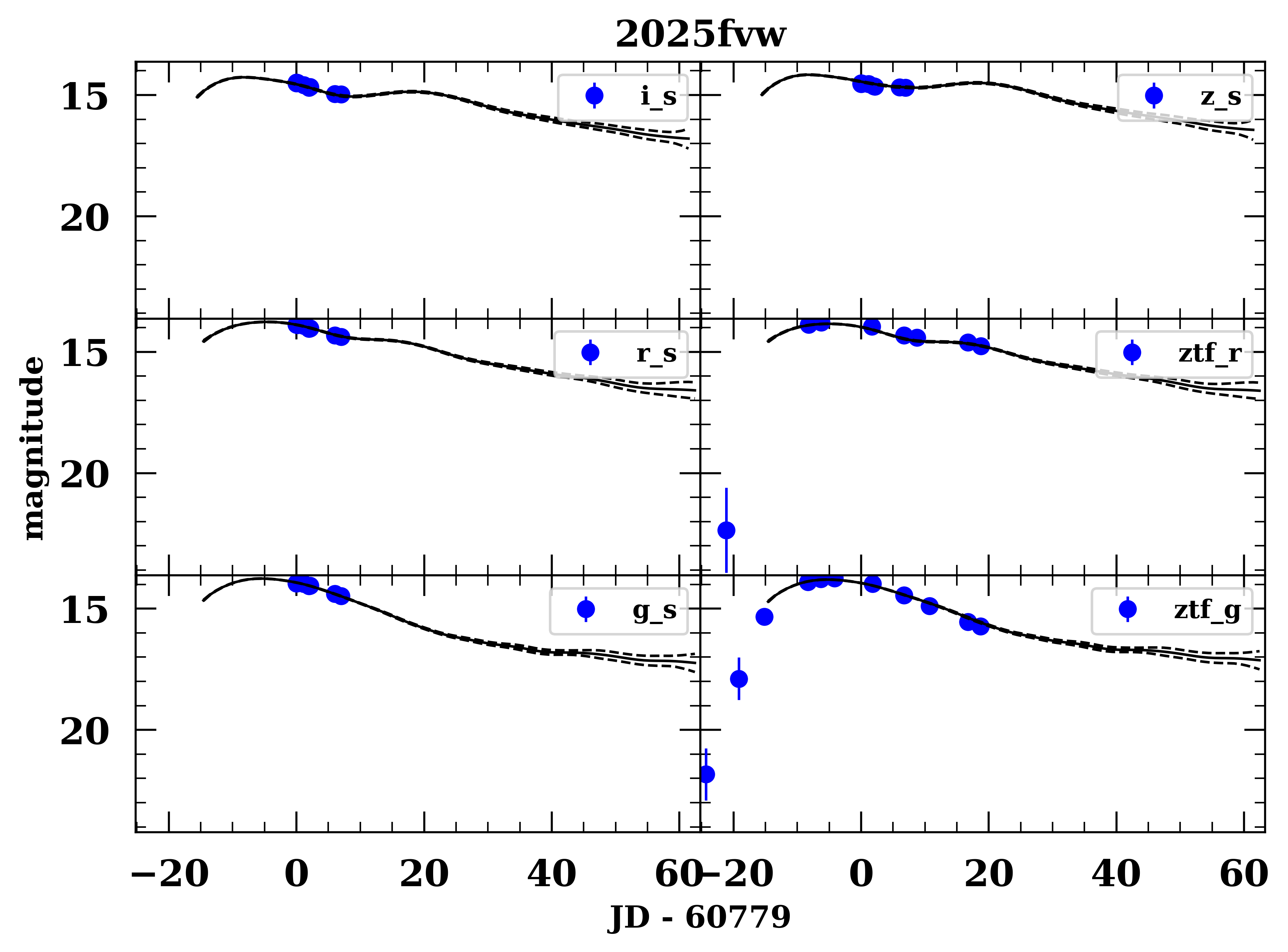}
\includegraphics[width=0.49\textwidth]{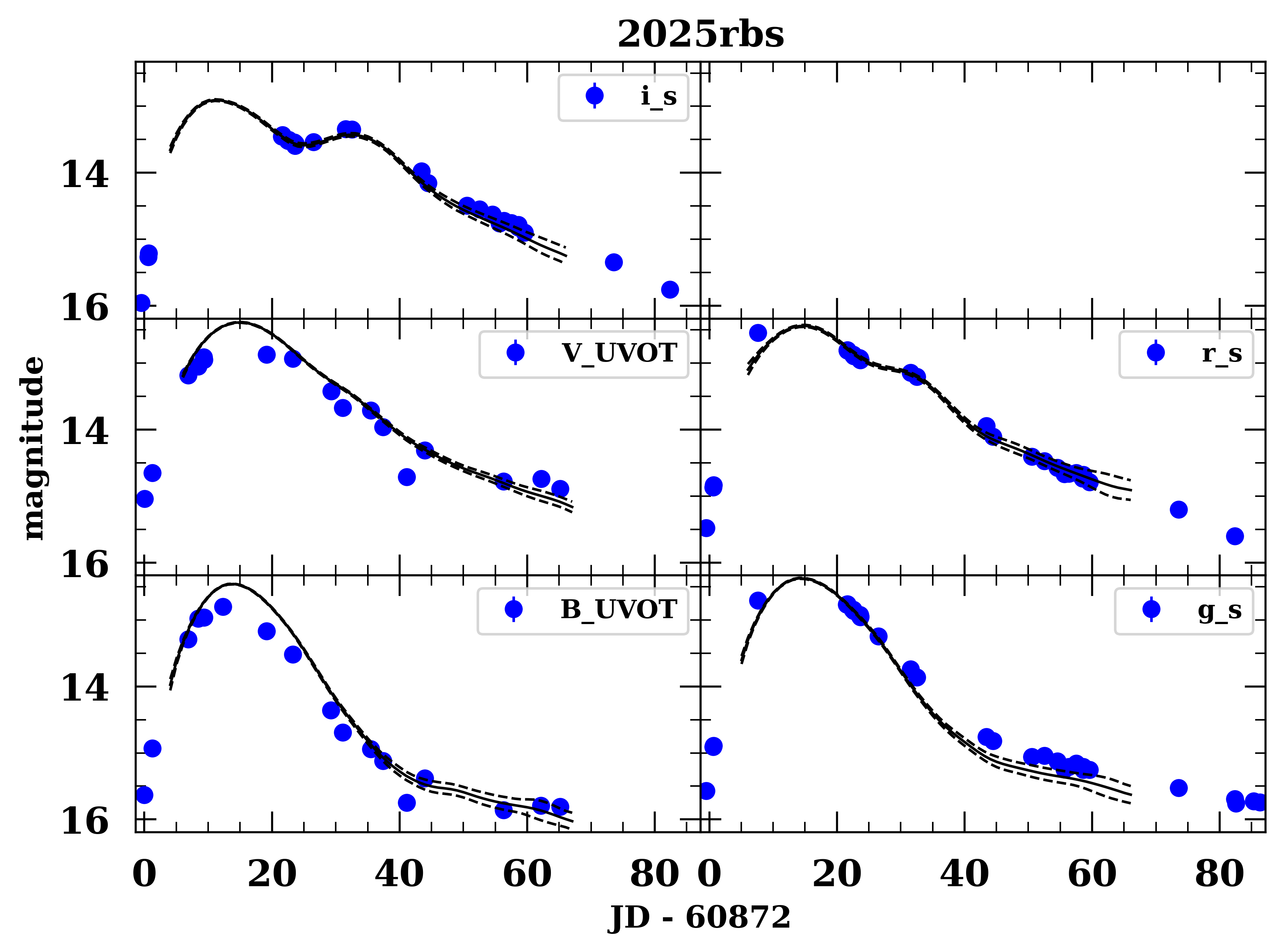}
\caption{\textit{Upper left}: \texttt{SNooPy} light-curve fitting for SN~2024inv. \textit{Upper right}: Same as left but for SN~2025fvw. \textit{Lower}: Same as above but for SN~2025rbs. The \texttt{EBV\_model} method is used in the fitting. The observations and best fit from \texttt{SNooPy} are shown in blue solid circles and black curves, respectively. The $g\_s$, $r\_s$, $i\_s$, $z\_s$ observations were obtained from the Lulin observatory. The $B\_UVOT$, $V\_UVOT$ observations were obtained from {\it Swift} UVOT. The $ztf\_g$, $ztf\_r$ observations were obtained from the Zwicky Transient Facility (ZTF) at the Palomar Observatory.}
\label{snoopy:fits}
\end{figure*}

Figure~\ref{snoopy:fits} shows the \texttt{SNooPy} fits for SN~2024inv, SN~2025fvw, and SN~2025rbs, obtained using the \texttt{EBV\_model} method. The photometric observations were obtained from the Lulin Observatory, {\it Swift}, and the Zwicky Transient Facility at Palomar Observatory.

\section{Results for \cahk}\label{cahk:app}

\begin{figure*}[htbp]
  \centering
  \includegraphics[width=0.79\textwidth]{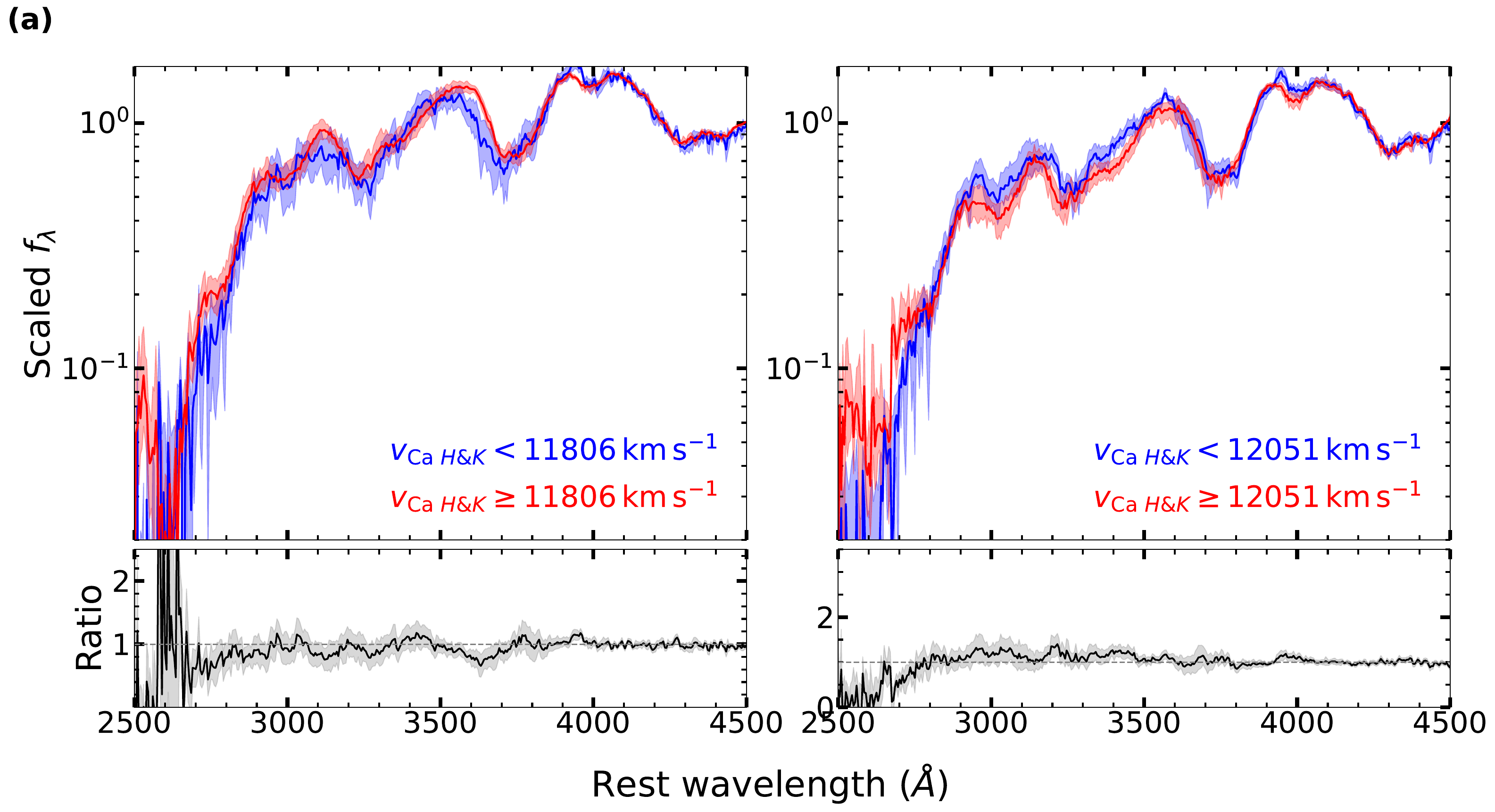}
  \includegraphics[width=0.79\textwidth]{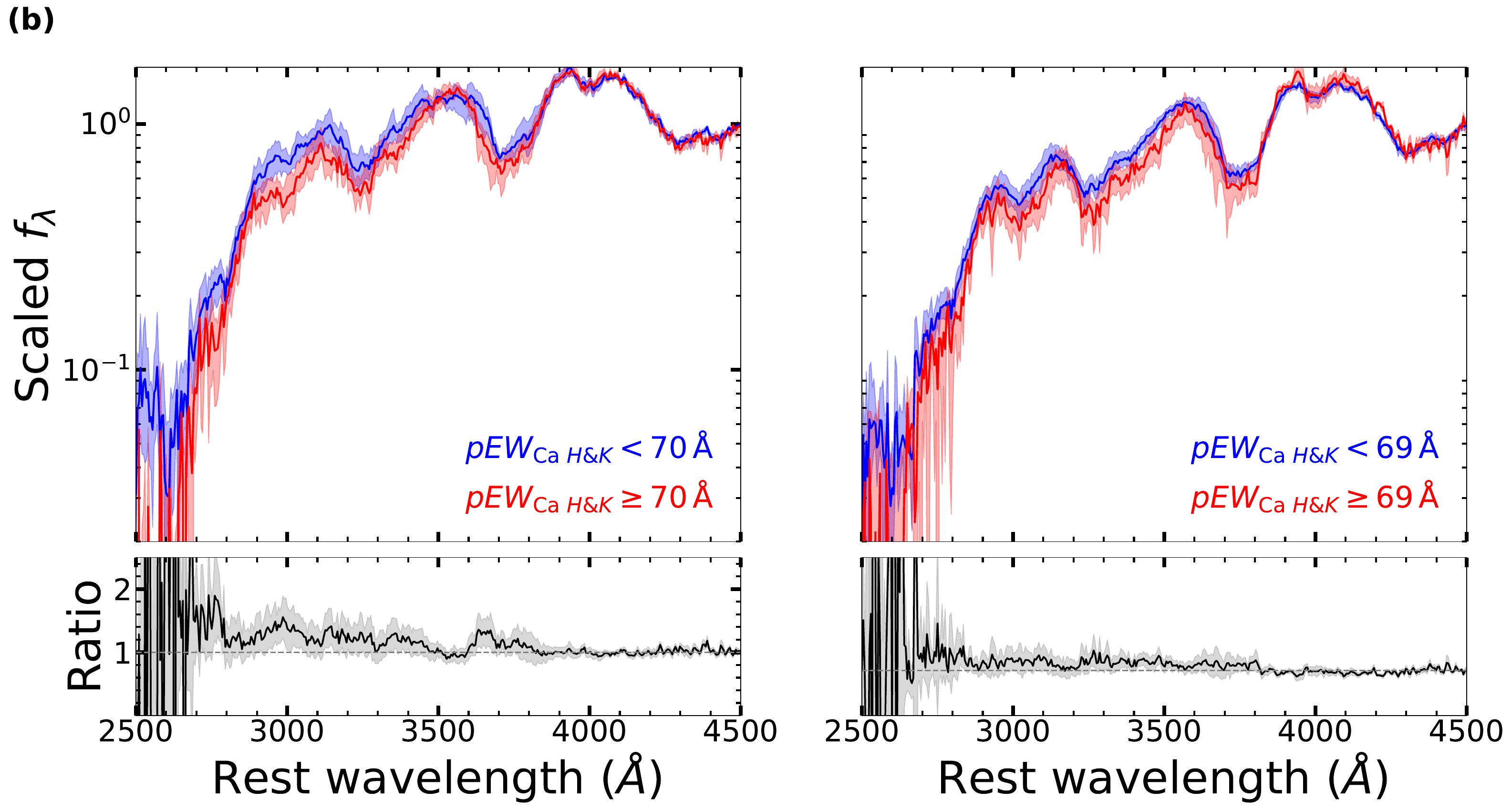}
  \caption{(a) \textit{Left}: Mean spectra with respect to \cahk\ photospheric velocity for the pre-maximum sample. \textit{Right}: Same as the left but for the near-peak sample. (b): Same as (a) but with respect to \cahk\ pEW.}
  \label{fig:cahk:mean}
\end{figure*}

\begin{figure*}[htbp]
  \centering
  \includegraphics[width=0.79\textwidth]{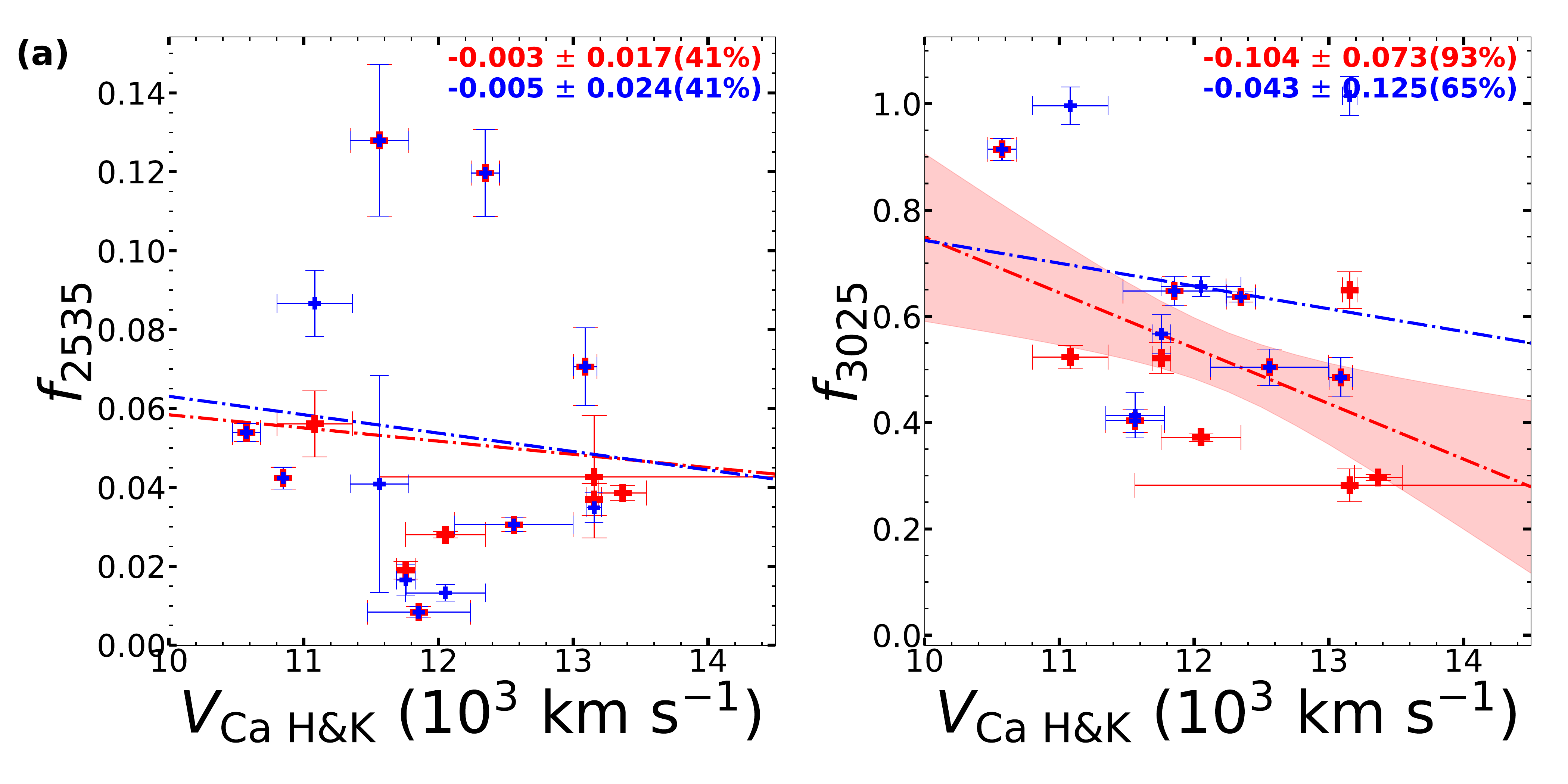}
  \includegraphics[width=0.79\textwidth]{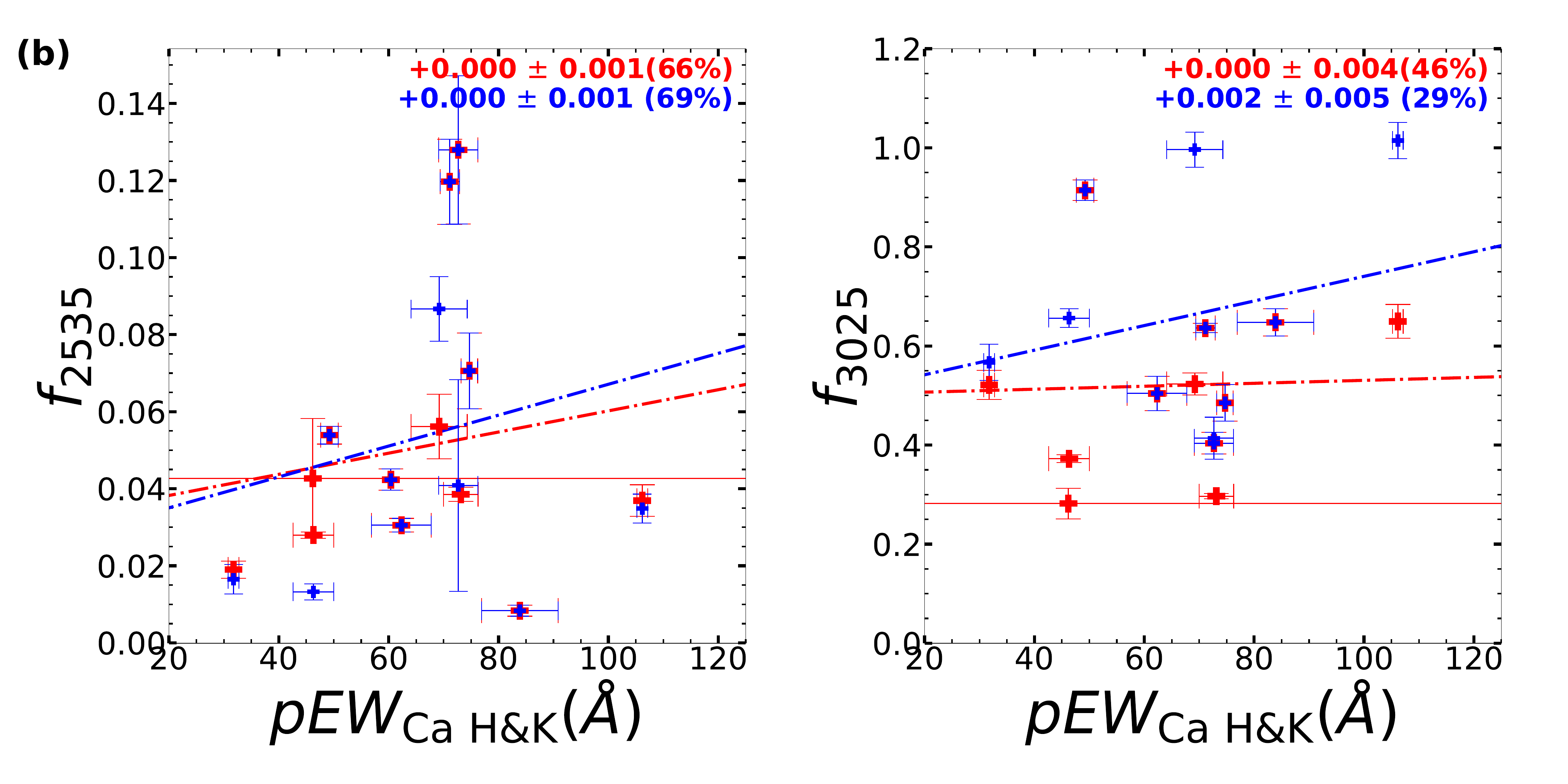}
  \caption{(a) \textit{Left}: The flux ratio $f_{2535}$ as a function of \cahk\ photospheric-velocity. \textit{Right}: Same as left but for the flux ratio $f_{3025}$. (b) Same as (a) but with respect to \cahk\ pEW.}
  \label{fig:cahk:ratio}
\end{figure*}

Figure~\ref{fig:cahk:mean} shows the mean UV spectra as a function of the photospheric velocity and pseudo-equivalent width (pEW) of the \cahk\ feature. For both parameters, we divide the pre-maximum and near-peak samples into two subgroups using their respective median values. We find that the difference between the two subgroups is qualitatively similar to that found for the \si\ velocity in Section~\ref{sec:mean}. The corresponding flux-ratio measurement in Figure~\ref{fig:cahk:ratio} shows a possible trend between $f_{3025}$ and \cahk\ velocity for the near-peak sample. We find no clear dependence on \cahk\ velocity or pEW in the pre-maximum sample, nor a clear trend with \cahk\ pEW in the near-peak sample.

\clearpage
\bibliography{sample7}{}
\bibliographystyle{aasjournalv7}



\end{document}

%% file: tardis_table.tex
\begin{table*}[ht]
\centering
\caption{Model ejecta composition for the base models and plasma configuration.}
\label{tab:tardis_ejecta}
\begin{tabular}{l l l}
\hline
\hline
\textit{Supernova}: \\
Luminosity requested ($\rm erg~s^{-1}$) & 7.67 $\times$ 10$^{42}$  & 9.8 $\times$ 10$^{42}$ \\
Time since explosion (in days) & 12  & 12 \\
$v_{inner} (km~s^{-1}$) & 12\,000  & 12\,000 \\
\hline
\textit{Composition}: \\
C  & 0.2  \\
O  & 0.77 \\
Mg & 1 $\times$ $10^{-4}$ \\
Si & 5.0 $\times$ $10^{-3}$ \\
S  & 2.0 $\times$ $10^{-2}$ \\
Ca & 1.5 $\times$ $10^{-3}$ \\
Ti & 3.0 $\times$ $10^{-4}$ \\
Cr & 3.0 $\times$ $10^{-4}$ \\
Fe & 5.0 $\times$ $10^{-4}$ \\
$\rm ^{56}Ni$ ($\rm t_{0} = 0~day$) & 5.0 $\times$ $10^{-3}$ \\
Isotope mass fractions at 12 days since explosion \\
$\rm ^{56}Ni$ & 0.00023 \\
$\rm ^{56}Co$ & 0.0035 \\
$\rm ^{56}Fe$ & 0.0013 \\
\hline
\hline
\textit{Plasma}: \\
electron scattering   & enabled \\
ionization            & nebular \\
excitation            & dilute-lte \\
radiative rates type  & dilute-blackbody \\
line interaction type & macroatom \\
\hline
\hline
\end{tabular}
\end{table*}